\documentclass[a4paper,11pt]{article}
\usepackage[margin=0.7in]{geometry}
\usepackage{amsmath}
\usepackage{amsfonts}
\usepackage{amssymb}
\usepackage{graphicx,caption}
\usepackage{epstopdf}
\usepackage{natbib}
\usepackage{enumerate}
\usepackage{float}
\usepackage[explicit]{titlesec}
\usepackage{lscape}
\pdfoutput=1

\usepackage{multirow}
\usepackage{tabularx,booktabs}
\usepackage{rotating}

\newlength{\verticalcompensationlength}
\newcounter{verticalcompensationrows}
\newcommand{\verticalcompensation}[1]{%
\setcounter{verticalcompensationrows}{#1}%
\addtocounter{verticalcompensationrows}{-1}%
\vspace*{-\value{verticalcompensationrows}\verticalcompensationlength}%
}
\newcommand{\multirowbt}[3]{%
\multirow{#1}{#2}{\verticalcompensation{#1}#3}%
}
\begin{document}

   \title{Puzzling out the coexistence of terrestrial planets and giant exoplanets. \\The 2/1 resonant periodic orbits}

   \author{Kyriaki I. Antoniadou and Anne-Sophie Libert\vspace{0.25cm}	\\
				\small{NaXys, Department of Mathematics, University of Namur, 8 Rempart de la Vierge, 5000 Namur, Belgium} \\  \small{kyriaki.antoniadou@unamur.be}}

\date{}
\maketitle

\begin{center}
The final publication is available at\\ https://doi.org/10.1051/0004-6361/201732058
\end{center}

\begin{abstract} 
Hundreds of giant planets have been discovered so far and the quest of exo-Earths in giant planet systems has become intriguing. In this work, we aim to address the question of the possible long-term coexistence of a terrestrial companion on an orbit interior to a giant planet, and explore the extent of the stability regions for both non-resonant and resonant configurations.
Our study focuses on the restricted three-body problem, where an inner terrestrial planet (massless body) moves under the gravitational attraction of a star and an outer massive planet on a circular or elliptic orbit. Using the Detrended Fast Lyapunov Indicator as a chaotic indicator, we constructed maps of dynamical stability by varying both the eccentricity of the outer giant planet and the semi-major axis of the inner terrestrial planet, and identify the boundaries of the stability domains. Guided by the computation of families of periodic orbits, the phase space is unravelled by meticulously chosen stable periodic orbits, which buttress the stability domains.
We provide all possible stability domains for coplanar symmetric configurations and show that a terrestrial planet, either in mean-motion resonance or not, can coexist with a giant planet, when the latter moves on either a circular or an (even highly) eccentric orbit.  New families of symmetric and asymmetric periodic orbits are presented for the 2/1 resonance. It is shown that an inner terrestrial planet can survive long time spans with a giant eccentric outer planet on resonant symmetric orbits, even when both orbits are highly eccentric. For 22 detected single-planet systems consisting of a giant planet with high eccentricity, we discuss the possible existence of a terrestrial planet. This study is particularly suitable for the research of companions among the detected systems with giant planets, and could assist with refining observational data.
 \end{abstract}

{\bf keywords}celestial mechanics --
planetary and satellites: dynamical evolution and stability -- minor planets, asteroids: general -- planetary systems -- methods: analytical -- methods: numerical

\section{Introduction}
Thus far an unrivalled number of exoplanets has been brought to light by missions or ground based telescopes\footnote{See e.g. {\em exoplanet.eu} \citep{enc} and {\em exoplanets.org} \citep{exorg}}. Given the surprisingly different exoplanetary configurations with respect to the solar system, among questions being raised regarding their formation and dynamical evolution, the scientific community has also been intrigued by the quest of “exo-Earths”. In particular, for the giant planetary systems discovered with the radial velocity (RV) method, the detection of Earth-like planets in these systems is currently elusive, due to observational limitations. For this reason, a theoretical study of the potential locations of Earth-like planets in these systems can prove to be a fruitful venture. 

A potentially habitable planet is a terrestrial planet within the habitable zone (HZ), meaning that it is within the region around a star where water in liquid form can be maintained on its surface \citep{kast}. The long-term stability of such a planetary orbit is a crucial factor for the biosphere to evolve. Many extrasolar systems consist of more than one planet, so that the orbit of an Earth-like planet evolves under the pertubations of its companion planets. These perturbations can weaken the long-term stability of the terrestrial orbit. Dynamical studies are thus essential to determine whether a given planet can remain stable for long time spans in the HZ. The sustainability of habitable terrestrial exoplanets, under the effect of a companion giant planet (whether in mean-motion resonance or not), is the central question of the present work.

The framework of our study is the three-body problem (TBP). It has been shown in the past that planets in mean-motion resonance (MMR) prompt the investigation of the dynamics. In a suitable frame of reference, the computation of the families of stable periodic orbits, which constitute the backbone of stability domains in phase space, can act as a diagnostic tool that ascertains information regarding the dynamical neighbourhood of exoplanets \citep[see e.g.][]{femibe06,hadj06,henli,av16,a16}. Moreover, families of periodic orbits can formulate the paths that drive the migration process of the planets trapped in MMR \citep[see e.g.][]{mebeaumich03,lee04,vat14,av17}.

The stability regions around the families of periodic orbits can be revealed by constructing maps of dynamical stability (DS-maps) with the use of a chaotic indicator. Different problems were tackled with this method. For instance, \citet{sandor07} realised a catalogue of DS-maps of hypothetical terrestrial planets in the HZ of 15 known exoplanetary systems with one giant. \citet{funk09} explored the stability of inclined terrestrial planets evolving in the HZ for four different configurations: terrestrial planets in (i) binary systems, (ii) with an inner gas giant, (iii) with an outer gas giant, and (iv) on a Trojan orbit. \citet{funkK} investigated the influence of the Kozai mechanism induced by an eccentric giant planet on the long-term stability of inclined Earth-like planets in the HZ.  

Here we have performed an exploration of the dynamical phase space of the coplanar restricted three body problem (RTBP) consisting of an inner terrestrial (massless) planet and an outer giant massive one, either trapped in interior MMR or evolving in non-resonant configurations. We showcase the influence of stable symmetric periodic orbits on the survival of such planetary systems.  Our work extends the study of \citet{sandor07} to a broader range of initial configurations, and is enhanced by an analytical study of the periodic orbits shaping the domains of stability for the 2/1 MMR. Our aim is to provide hints as to the long-term stability and possible existence of both circular (Earth-like) and quite eccentric inner terrestrial planets when evolving in systems with an outer giant planet regardless of its eccentricity. 

The paper is organised as follows. In Sect. \ref{model}, we present the model and discuss the basics with regards to the bifurcation and continuation of the symmetric and asymmetric planar periodic orbits and their link with MMRs. The chaotic indicator used in the DS-maps is also described. In Sect. \ref{maps}, we show a series of DS-maps, when all possible symmetric configurations are taken into account as the semi-major axes ratio varies along with the eccentricity of the outer giant, for four different values of the eccentricity of the inner terrestrial planet. In order to showcase that the regular domains in phase space do encompass the stable periodic orbits, we have focused on 2/1 MMR. In Sect. \ref{21}, we provide the families of periodic orbits in the circular and elliptic restricted TBPs for 2/1 MMR. 
We justify the islands of stability that appear in the dynamical neighbourhood of 2/1 MMR in the DS-maps of Sect. \ref{maps}, by computing additional DS-maps initiated by specific stable periodic orbits on different planes, so that a global view of the phase space is provided. In Sect. \ref{application}, we apply our results to RV-detected single-giant planet systems by showcasing the dynamical vicinity of possibly existing terrestrial planet. Finally, our conclusions on such a coexistence are given in Sect. \ref{con}. 

\section{Model set-up}\label{model}
We considered two planets, a massless planet ($m_1=0$), named terrestrial planet here, and a giant one of mass $m_2$, both of which revolve around a star of mass $m_0$. The massless body moves under the gravitational attraction of $m_0$ and $m_2$ (without influencing their motion), as the latter describes circular (CRTBP) or elliptic (ERTBP) orbits around their common centre of mass in a suitable frame of reference $Oxy$ \citep{hach75}. The parameter of the RTBP is $\mu=\frac{m_2}{m_0+m_2}$. The planetary orbits correspond to Keplerian ellipses in the inertial frame described by heliocentric osculating elements, namely the semi-major axes $a_i$, the eccentricities $e_i$ and the longitudes of pericentre $\varpi_i$. For the position of the planets on the osculating ellipse, we considered the mean anomalies $M_i$. Subscripts 1 and 2 refer to the inner and the outer planet, respectively.  We normalised the total mass of the bodies to unity, i.e. $m_0+m_1+m_2=1$. Hence, since $m_1=0$, the mass of the star is $m_0=1-m_2$ with $m_2=0.001$. Additionally, for the gravitational constant $G$, we used the normalisation $G=1$. 

Without loss of generality, the initial semi-major axis of the outer planet is always set to $a_2(0)=1.0$. Thus, we study interior MMRs, meaning that the mean-motion ratio is rational defined as $\frac{n_2}{n_1}=\left(\frac{a_1}{a_2}\right)^{-3/2}\approx\frac{p+q}{p}$, where $p, q \in \mathbb{Z}^*$ and $q$ is the order of the resonance. 

For the computation of DS-maps, we focused on coplanar symmetric periodic orbits, where the apsidal difference was fixed to $\Delta\varpi=\varpi_2-\varpi_1=0$ or $\pi$. Four values of the eccentricity of the inner terrestrial planet, $e_1$, are considered here: 0.02, 0.1, 0.3 and 0.5. We report all possible symmetric configurations for the semi-major axes ratios, $a_1/a_2$ with $a_2=1.0$ varying between $0.1$ and $0.8$, and for all values of the eccentricity of the outer giant planet, i.e. $e_2 \in [0,1]$. To explain dynamically the stable regions observed in the DS-maps, families of symmetric periodic orbits in the ERTBP were computed for the 2/1 MMR. We note that families of asymmetric periodic orbits are also investigated for reasons of completeness. 

We performed the numerical integrations of the equations of motion with the Bulirsch-Stoer integrator. Needless to say, a broad exploration of regular orbits in the phase space of such systems requires extensive numerical integrations. 
Computation of the linearly stable\footnote{Linear stability is derived through the computation of the eigenvalues of the system. The periodic orbits are considered as stable if and only if all of the eigenvalues are lying on the unit circle. See also Sect. \ref{pos}.} periodic solutions that guide our exploration of the precise boundaries of each stable domain, is further discussed in Sect. \ref{pos}. The long-term stability of trajectories within broad regions in phase space is also concluded by the computation of the Detrended Fast Lyapunov Indicator, as set out in Sect. \ref{DFLI}.

\subsection{Periodic orbits and mean-motion resonances}\label{pos}

Let us consider the rotating frame of reference $Oxy$, whose origin coincides with the centre of mass of the star and the giant planet (primaries) and its $x$-axis always contains them. The positions of the planet and the massless body are $(x',0)$ and (x,y), respectively. Then, when $\dot x'=0,$ a periodic orbit is defined as $\textbf{Q}(0) = \textbf{Q}(T)$, where $\textbf{Q}(t)$ is a set of positions and velocities of both the giant planet and the terrestrial one and $T$ is the orbit's period, which satisfies $t = kT$, with $k \ge 1$ being an integer. Given the Lagrangian of the system and the respective equations of motion, the system remains invariant under certain periodicity conditions, which determine whether the periodic orbit is symmetric or asymmetric \citep[see e.g.][]{avk11}. 

In order to classify the orbits we took into account the fundamental symmetry, $\Sigma$, \citep{hen97}
\begin{equation} 
\Sigma: (t,x,y) \rightarrow (-t,x,-y).
\end{equation}

A periodic orbit coincides with a fixed or periodic point on a Poincar\'e surface of section, say at $y=0$ with $\dot y>0$. A periodic orbit is symmetric when it has two perpendicular crossings with the $Ox$-axis, i.e. $y(T)=y(0)=0$ and $\dot{x}(T)=\dot{x}(0)=0$.

The symmetric periodic orbits with respect to the $x$-axis in the ERTBP are represented by a point in the three dimensional phase space 
\begin{equation}\begin{array}{lll}
        x'(T)=x'(0),& x(T)=x(0), & \dot{y}(T)=\dot{y}(0), 
\end{array}\end{equation}
where  $x'$ denotes the position of the giant planet, while $x$ and $\dot y$ the position and velocity of the terrestrial planet in the rotating frame. 

If the orbit starts on the plane $(x,y)$, the asymmetric periodic orbits (provided that $\dot{x'}(T)=\dot{x'}(0)=0$) are represented in the ERTBP by a point in the five-dimensional space of initial conditions
\begin{equation}\begin{array}{lll}
        x'(T)=x'(0),& x(T)=x(0),&y(T)=y(0),\\
        \dot{x}(T)=\dot{x}(0),&\dot{y}(T)=\dot{y}(0).
\end{array}\end{equation}
If the orbit does not start perpendicularly from the $Ox$-axis, the asymmetric periodic orbits (provided that $y(T)=y(0)=0$) are represented in the ERTBP by the conditions
\begin{equation}\begin{array}{lll}
        x'(T)=x'(0),& \dot {x'}(T)=\dot {x'}(0)\neq0,&x(T)=x(0),\\
                 \dot{x}(T)=\dot{x}(0),&\dot{y}(T)=\dot{y}(0).
\end{array}\end{equation}
    
In the CRTBP, $x'$ is constant, yielded by the normalisation adopted for the system and in our computations is taken equal to $1-\mu$. Thus, a symmetric periodic orbit with respect to the $x$-axis is defined as a point in the space of initial conditions
\begin{equation}\begin{array}{ll}
x(T)=x(0), & \dot{y}(T)=\dot{y}(0). 
\end{array}\end{equation}

An asymmetric periodic orbit in the CRTBP, when the reference is $\dot{x'}(T)=\dot{x'}(0)=0$ is defined as
\begin{equation}\begin{array}{llll}
 x(T)=x(0),&y(T)=y(0), & \dot{x}(T)=\dot{x}(0),&\dot{y}(T)=\dot{y}(0), 
\end{array}\end{equation}
or as
\begin{equation}\begin{array}{llll}
 \dot {x'}(T)=\dot {x'}(0)\neq0,&x(T)=x(0),&\dot{x}(T)=\dot{x}(0),&\dot{y}(T)=\dot{y}(0), 
\end{array}\end{equation}
when the reference is $y(T)=y(0)=0$.

Then, via the mono-parametric continuation \citep[see e.g.][]{hen97,hadj06}, smooth curves known as characteristic curves, or families of periodic orbits are derived linking those fixed points. The linear stability analysis \citep{Broucke1969,marchal90,hadjbook06} can classify the periodic orbits as stable or unstable and based on this attribute we are able to foretell the evolution of celestial bodies hosted in their neighbourhood. 

In principle, the families of periodic orbits are generated by bifurcation points and continued according to specific continuation schemes. \citet{av12} studied the vertical stability of the planar CRTBP in 2/1 and 1/2 MMRs and general TBP (GTBP) in 2/1 MMR, where both bodies are massive and the mass-ratio $\rho=\frac{m_2}{m_1}$ changes the dynamics, so there is no distinction between interior ($\rho\rightarrow \infty$) and exterior ($\rho\rightarrow 0$) MMRs. They showed the bifurcation and continuation of spatial symmetric periodic orbits in 2/1 MMR for the 3D-GTBP starting either from the 2D-GTBP or the 3D-CRTBP. The generation of families among the planar problems (CRTBP, ERTBP and GTBP) has been shown by e.g. \citet{hadj75}, \citet{vkh09} and \citet{avk11}. However, there exist families that do not bifurcate from periodic orbits and are isolated. For instance, \citet{av12} showed examples of spatial isolated symmetric families in 2/1 MMR in 3D-GTBP resulting by foldings of the families as the mass of the inner or outer body increases from zero and  \citet{voyhadj05}, \citet{vkh09} and \citet{av16} presented planar asymmetric isolated families in 2/1 MMR in GTBP resulting by collision bifurcations as $\rho=\frac{m_2}{m_1}$ changes.

A family of periodic orbits is either circular or elliptic.  The circular family, along which the mean-motion ratio, $\frac{n_2}{n_1}$, varies, consists of almost circular symmetric periodic orbits. At ratios at which we have an MMR, the circular family either continues smoothly with periodic orbits of increasing eccentricities (when $q=1$), or exhibits bifurcation points that generate new elliptic families (when $q\neq 1$). These elliptic families, along which the ratio $\frac{n_2}{n_1}$ remains close to its rational value for the RTBP, may be continued up to high eccentricities. Thus, they are resonant and each resonant periodic orbit they consist of indicates the exact position of the respective MMR in phase space at a particular energy level. 

The validity of an averaged Hamiltonian can be checked by comparing the periodic orbits of the RTBP with the corresponding fixed points (or stationary solutions) that depend on the resonant angles
\begin{equation}\begin{array}{l}
\theta_1=p\lambda_1-(p+q)\lambda_2+q\varpi_1, \\
\theta_2=p\lambda_1-(p+q)\lambda_2+q\varpi_2. \\
\end{array}
\end{equation}

The stationary solutions where $\dot{\theta_i}=0$ ($i=1,2$) are named as apsidal corotation resonances (ACRs e.g. \citet{femibe06} and \citet{mbf06}). Hence, an ACR corresponds to a periodic orbit in the rotating frame. In the neighbourhood of a stable periodic orbit, both the resonant angles and the apsidal difference, $\Delta\varpi=( \theta_2-\theta_1)/q$, librate about 0 or $\pi$ if the orbit is symmetric or around other angles if the orbit is asymmetric. By letting the resonant bodies be aligned ($\Delta\varpi$=0) or anti-aligned ($\Delta\varpi=\pi$) and by also considering the $M_i$ , we obtain four symmetric configurations given by $(\theta_1,\theta_2) = (0,0), (0,\pi), (\pi,0),$ and $(\pi,\pi)$. Along the families of asymmetric periodic orbits the angles $\Delta\varpi=\varpi_2-\varpi_1$ (measure of geometric asymmetry) and $\Delta M=M_2-M_1$ (measure of dynamic asymmetry) vary and we also have the mirror image at $\Delta\varpi'=2\pi-\Delta\varpi$ and $\Delta M'=2\pi-\Delta M$, as a result of $\Sigma$.

\subsection{Maps of dynamical stability}\label{DFLI}
It is widely known that in Hamiltonian systems, such as the planar ERTBP which has three degrees of freedom, the motion can be regular or chaotic. In domains of phase space surrounded by stable periodic orbits and thus, populated by invariant tori, the motion is regular and quasi-periodic according to KAM theory \citep{arnold}; \citep[see also e.g.][]{berry,cont}. Therefore, the long-term stability of the planetary systems residing therein is guaranteed, as Arnol'd diffusion is expected to be excessively slow. The regions around unstable periodic orbits exhibit weak or broad homoclinic chaos and the motion is irregular. Hence, the planetary systems in such a vicinity are destabilised and a collision or escape is probable. Weakly chaotic orbits can be considered as stable from a physical point of view, since the initial configuration of the planets does not change significantly over time.

Chaotic or regular evolution in the neighbourhood of periodic orbits can, in general, be revealed by computing chaotic indices \citep{sandor07}. Here we have used the {\em Detrended} Fast Lyapunov Indicator (DFLI) \citep{froe97,voyatzis08}, which is fast and reliable in distinguishing chaos from order, and is defined as
\begin{equation}                
DFLI(t)=log \left ( \frac{1}{t}\max\{|\mathbf{\eta_1}(t)|,|\mathbf{\eta_2}(t)|\} \right ),
\end{equation}
where $\eta_i$ are the initially orthogonal deviation vectors computed after numerical integration of the variational equations. For a regular orbit DFLI remains almost constant over time, while it increases exponentially when irregular orbits are traced.

We broadly computed DS-maps in order for the multi-dimensional phase space structure to be unravelled. By choosing suitable planes we created dense grids of initial conditions, while keeping the rest of the parameters fixed. Then, based on the output of the chaotic indicator we coloured each condition. Throughout this study, dark coloured regions represent the domains where the motion is regular, while pale coloured ones correspond to regions where chaoticity is traced. The white colour showcases the points where the numerical integration has failed at $t<t_{max}$, due to very close encounters, which, in turn, resulted in very small integration step. We have chosen $t_{max}=250 Ky$, which has been proved by numerical tests more than adequate to the efficient distinguishability of motion. For irregular orbits, we stopped the numerical integration, when $DFLI(t)>10^{30}$ and classified the orbit as chaotic. 

\begin{figure}[!h]
\centering
\resizebox{0.4\hsize}{!}{\includegraphics{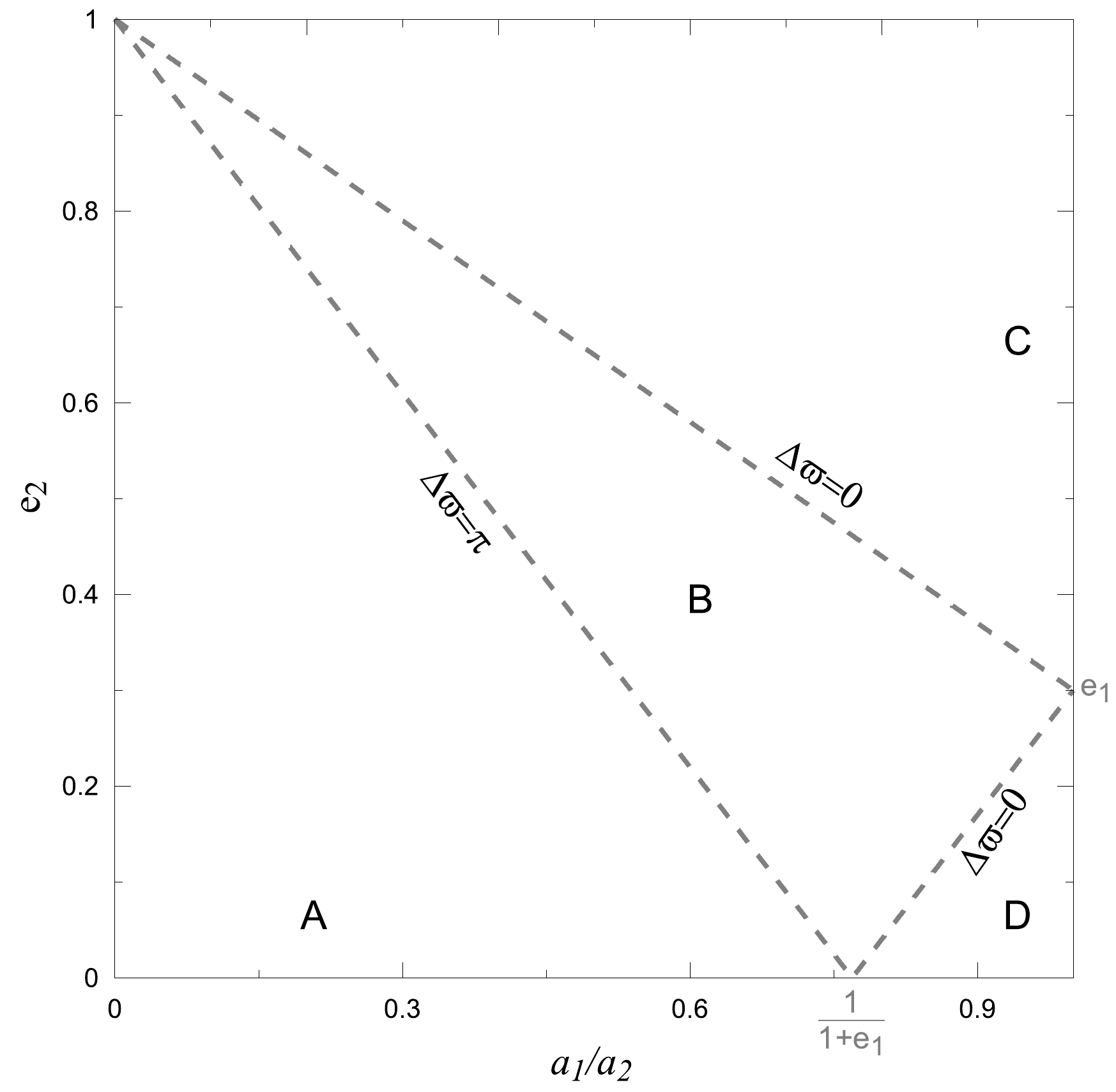}}
\caption{Collision lines for interior MMRs, when $e_1=0.3$ and $a_2=1.0$ that separate the plane $(a_1 / a_2,e_2)$ in four different domains denoted by capital letters. Their intersection with the axes is also shown.}
\label{colall}
\end{figure}

To unravel the possible symmetric configurations for all possible locations of the terrestrial planet, regardless of the eccentricity of the companion giant planet, the DS-maps are computed in the plane $(a_1/a_2,e_2)$. Although the MMRs can offer a phase protection mechanism and even highly eccentric orbits can survive collisions \citep{av16}, two coplanar Keplerian orbits may, in general, intersect if the criterion
\begin{equation}
a_1^2(1-e_1^2)+a_2^2(1-e_2^2)-2a_1 a_2 (1-e_1 e_2 \cos\Delta \varpi) \leq 0
\label{ColCri}
\end{equation}
holds \citep{kho99}. If we assume a constant eccentricity value $e_1$ and set $\Delta\varpi$ either equal to $0$ or $\pi$, it is trivial to yield the collision lines. In Fig. \ref{colall}, we show that the collision lines separate the plane $(a_1/a_2,e_2)$ in four different areas. In $A$, the planetary orbits do not intersect for any value of the apsidal difference, $\Delta\varpi$. In $B$, only anti-aligned orbits (or orbits with $\Delta\varpi$ precessing about $\pi$ and satisfying Eq. (\ref{ColCri})) intersect. In $C$ and $D$, the orbits always intersect.

\begin{figure*}[!ht]
\centering
\includegraphics[width=17.5cm]{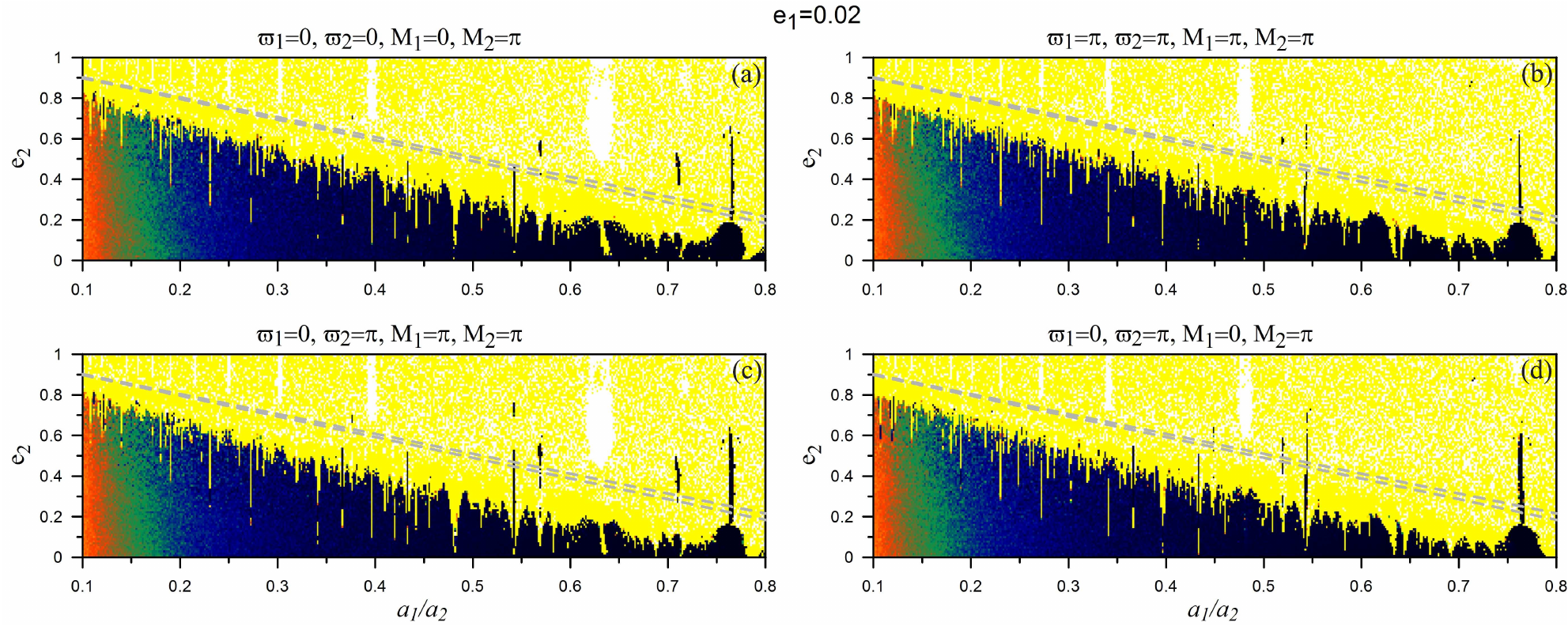}\\
\includegraphics[width=3.cm,height=0.6cm]{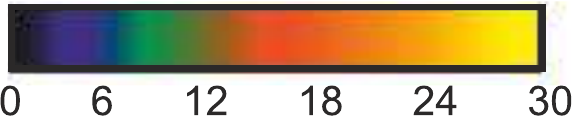} 
\caption{DS-maps on the plane ($a_1/a_2,e_2$) when $e_1=0.2$. The values of the orbital elements that remained fixed throughout the computation of the DS-maps are noted above each panel. The dashed grey lines indicate the collision lines between the planets. The coloured bar corresponds to the logarithmic values of DFLI; dark (pale) colours showcase regular (irregular) evolution of the orbits, while white colour depicts the very close encounters and the failure of numerical integration at $t<250 Ky$.}
\label{002}
\end{figure*}

\begin{figure*}[!ht]
\centering
\includegraphics[width=17.5cm]{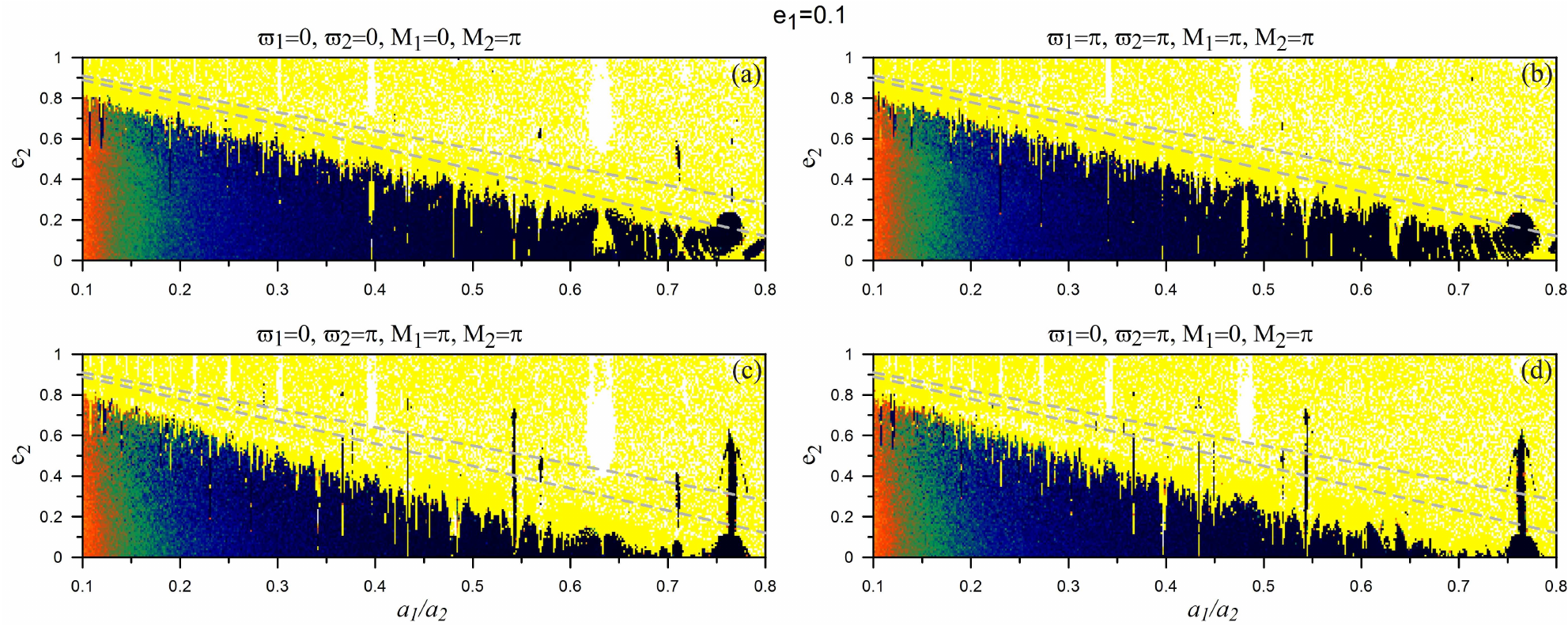}
\caption{DS-maps on the plane ($a_1/a_2,e_2$) when $e_1=0.1$. Colours and lines as in Fig. \ref{002}.}
\label{01}
\end{figure*}

\begin{figure*}[!ht]
\centering
\includegraphics[width=17.5cm]{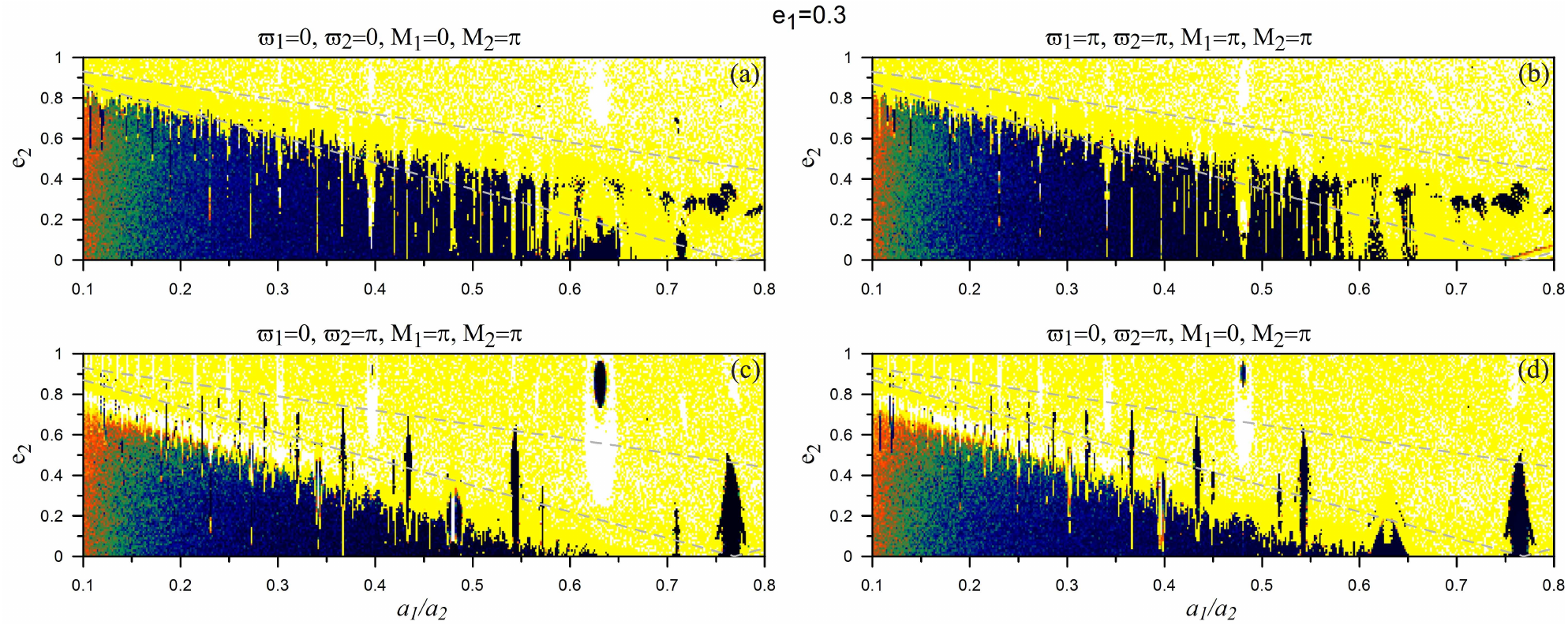}
\caption{DS-maps on the plane ($a_1/a_2,e_2$) when $e_1=0.3$. Colours and lines as in Fig. \ref{002}.}
\label{03}
\end{figure*}

\begin{figure*}[!ht]
\centering
\includegraphics[width=17.5cm]{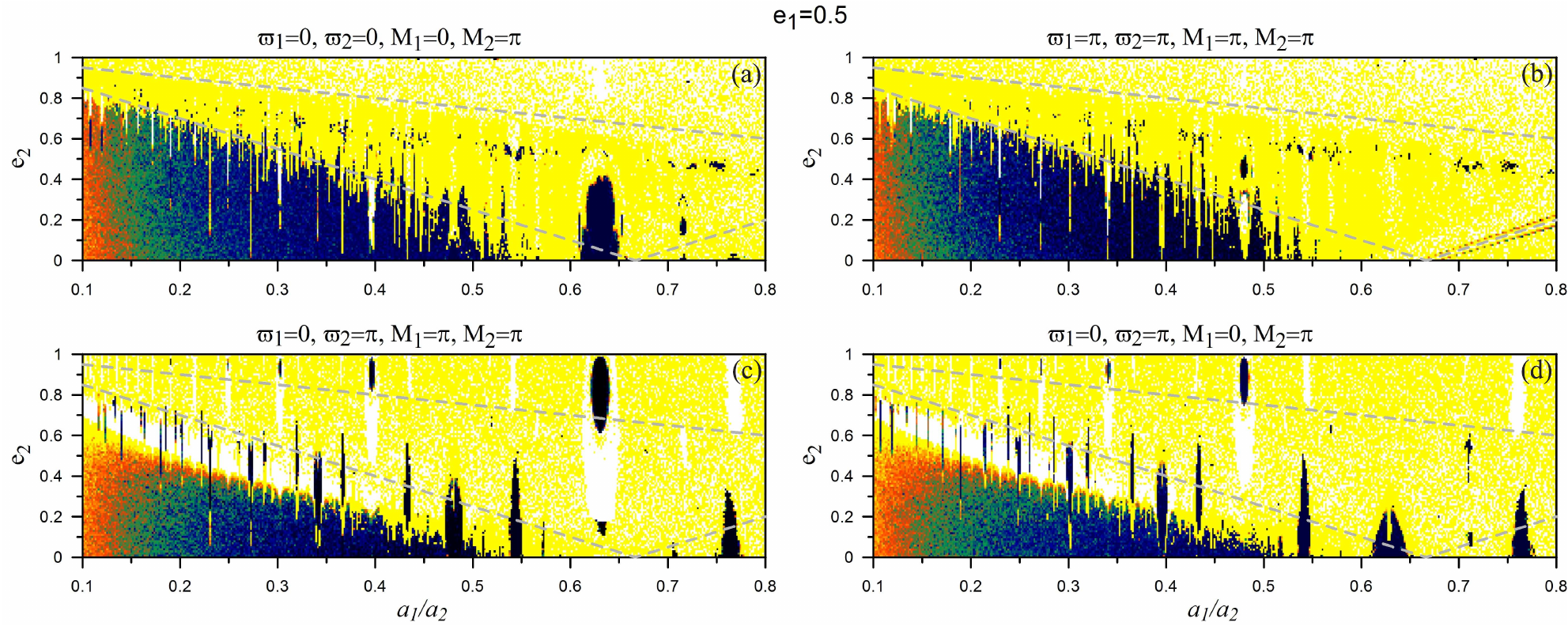}
\caption{DS-maps on the plane ($a_1/a_2,e_2$) when $e_1=0.5$. Colours and lines as in Fig. \ref{002}.}
\label{05}
\end{figure*}

\section{Unravelling the existence of terrestrial planets}\label{maps}

Hereafter, we present DS-maps on the plane ($a_1/a_2,e_2$), where $a_2=1.0$ by keeping $e_1$ fixed and equal to $0.02$, $0.1$, $0.3$, and $0.5$. As for the rest of the orbital elements and as already stated, when it comes to symmetric periodic orbits, four different symmetric configurations exist, provided by the resonant angles. Thus, each time (for each fixed value of $e_1$), we present four DS-maps, where the respective values of $M_i$ and $\varpi_i$ are fixed. In this way we obtain a global view of the phase space of coplanar symmetric planetary configurations.
  
We herein begin a thorough study of the phase space around 3/2, 2/1, 5/2, 3/1, 4/1, and 5/1 MMRs, while the semi-major axes ratio ranges between [0.1,0.8]. In this section, we provide information about the resonant and non-resonant stable motion via those DS-maps, where we overplot the collision lines according to each value of $e_1$. In Sect. \ref{21}, we explain the reason why those islands of stability exist for 2/1 MMR guided by the stable periodic orbits. In a forthcoming work \citep{spis}, such a link with periodic orbits shall be provided for each of the above-mentioned MMRs.

The DS-maps associated with $e_1=0.02$, $0.1$, $0.3,$ and $0.5$ are given in Figs. \ref{002}-\ref{05}, respectively. We note a similar trend in all the DS-maps. Following the collision lines, the larger the semi-major axes ratio of the two bodies, the smaller the eccentricity of the giant planet compatible with the existence of an inner terrestrial planet on stable orbit (see the shape of the broad region of regular orbits). However, at specific MMRs, interesting regular regions can additionally be observed at large values of the eccentricity $e_2$. In the following, we describe these specific attributes showcased in the DS-maps.    

To begin with, for low to moderate values of the eccentricity of the giant planet, $e_2$, we observed (regardless of $\Delta\varpi$ or $e_1$) distinct regions of stability (dark coloured) spanning from one domain to another being crossed by the collision lines (grey dashed curves). These islands, or rather tongues, emanating from domains $A$ or $D$ at semi-major axes ratios corresponding to MMRs, are built about stable symmetric periodic orbits. Therefore, they are not broken by the collision lines, as the MMRs offer a phase-protection mechanism and even when the orbits do intersect, the close encounters are avoided \citep[e.g.][]{morby}. On the other hand, for highly eccentric giants, meaning those above the collision lines in domain $C$, only distinct islands of stability exist, which are built about stable periodic orbits and thus, again, close encounters are not in effect. These will be justified through the linear stability of the periodic orbits in Sect. \ref{21}, for the 2/1 MMR.

We note that, at all those stability domains, both the respective resonant angles and the apsidal difference librate about $0$ or $\pi$. This feature justifies the capture in MMR and the evolution in such a stable symmetric configuration. Being close to a periodic orbit (or ACR, for apsidal corotations) results in oscillations of the orbital elements and angles of small amplitude. On the contrary, moving further away gives rise to oscillations of large amplitude.

Moreover, for the non-resonant regular orbits of domains $A$ and $B$, the apsidal resonance \citep{murray,morby}, where only the apsidal difference oscillates about 0, while the resonant angles rotate, is apparent. The orbits are thus, protected from close encounters \citep{malho02}\footnote{We note that, in case of apsidal resonance, the amplitude of one proper frequency becomes zero (or near zero), so both eccentricities are dominated by the other proper frequency, i.e. they precess with the same frequency. In true resonance (like the MMR), two proper frequencies of a system become commensurable.}.

As the terrestrial planet becomes more eccentric ($e_1\rightarrow 0.5$), we observe that the broad region of regular orbits shrinks towards lower values of $e_2$. Simultaneously, more distinct islands of stability and tongues appear. 

More precisely, in Fig. \ref{002}, the possible coexistence of an exo-Earth with a giant planet is investigated when $e_1=0.02$. Apart from the distinct tongues emerging at MMRs and the islands of stability in domain $C$, the long-term stability of the non-resonant systems in domains $A$ and $B$ is obtained through the apsidal resonance. The same holds for Fig. \ref{01}, where $e_1=0.1$. 

In Figs. \ref{03} and \ref{05}, for $e_1=0.3$ and 0.5, respectively (apart from the tongues extending from the domain $A$ to the domain $C$ and the islands in domain $C)$ the secondary resonance \citep{momo93} is noteworthy for the islands appearing at low eccentricities of the giant ($e_2<0.2$). Within these regions $\theta_1$ librates about 0 or $\pi$ (depending on the configuration of the periodic orbit (MMR) it is associated with), while $\theta_2$ and $\Delta\varpi$ rotate. More specifically, if we consider the average frequencies $\textit{f}_{2R}$ and $\textit{f}_{1L}$ of $\theta_2$-rotation and $\theta_1$-libration, respectively,  we can observe the existence of resonances $\textit{f}_{2R}/\textit{f}_{1L}=g_2/g_1$, where $g_i \in \mathbb{Z}^*$, known as secondary resonances.

\section{2/1 resonant periodic orbits}\label{21}

The goal of this section is to identify the 2/1 resonant periodic orbits that generate the tongues and islands of stability observed in the DS-maps. However, for reasons of completeness, apart from the families of symmetric periodic orbits in the CRTBP and ERTBP responsible for the regular domains of the DS-maps, we also present families of asymmetric periodic orbits, which exist only in the ERTBP.

\subsection{Symmetric periodic orbits}
\begin{figure}[!h]
\centering
\resizebox{0.4\hsize}{!}{\includegraphics{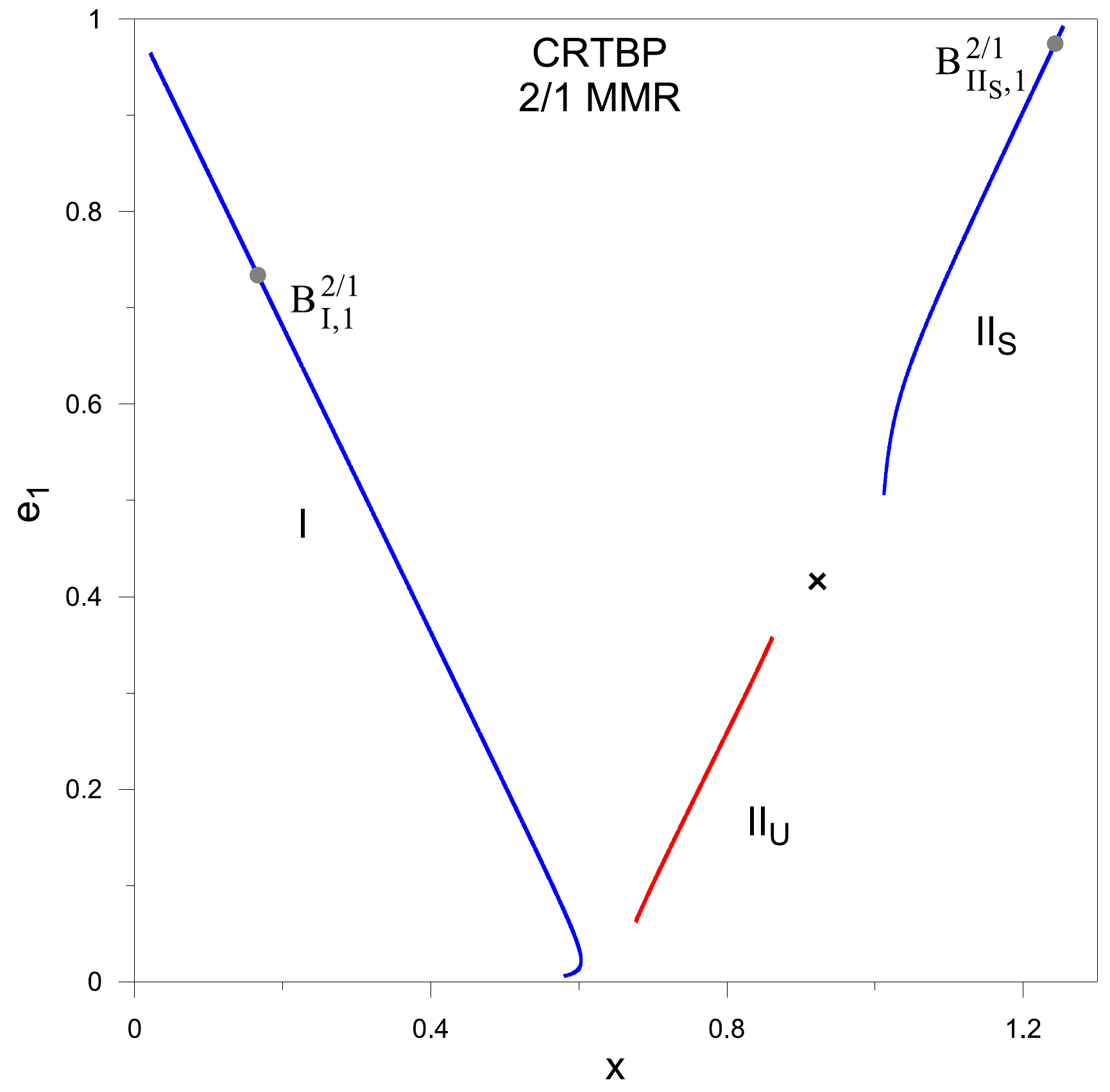}}
\caption{Families of periodic orbits in 2/1 MMR of the CRTBP when $m_2=0.001$ projected on the plane $(x,e_1)$. The family $I$ consists of periodic orbits which are aligned ($\Delta\varpi=0$), whereas along $II_U$ and $II_S$ the periodic orbits are anti-aligned ($\Delta\varpi=\pi$). The collision between the orbits is denoted by the cross symbol. Blue (red) stands for stable (unstable) periodic orbits. The bifurcation points from the CRTBP to the ERTBP are also shown.}
\label{21crtbp}
\end{figure}

In Fig. \ref{21crtbp}, we present the families of periodic orbits in the 2/1 MMR of the CRTBP originating from $x\approx 0.62996$. More specifically, along the circular family, whose orbits are circular and symmetric, the semi-major axis of the massless body (recall that $a_2=1.0$) coincides with the radius of its orbit and thus, the coordinate $x$ in the rotating frame. At points where $T=2\pi (a_1^{-3/2}-1)^{-1}$ we obtain bifurcation points that generate symmetric periodic orbits (and then though mono-parametric continuation families of periodic orbits) to the CRTBP. At first order MMRs where $q=1$, like the 2/1, as mentioned in Sect. \ref{pos}, the circular family exhibits a gap (see e.g. Fig. \ref{21crtbp} where the families $I$ and $II_U$ do not bifurcate smoothly from one point), as we switch on the mass of the giant planet from zero and go from the unperturbed to the perturbed problem. Due to the Poincar\'e-Birkhoff theorem, only a finite number of families survives. This number is usually two, one being stable ($I$) and one unstable ($II_U$).  Since the massless body, say the terrestrial planet, is now allowed to describe elliptic orbits and either be located at pericentre or apocentre, we present those families on the plane $(x,e_1)$. The family $I$ consists of aligned periodic orbits and the families $II_U$ and $II_S$ of anti-aligned ones.

\begin{figure}[!h]
\centering
\resizebox{0.4\hsize}{!}{\includegraphics{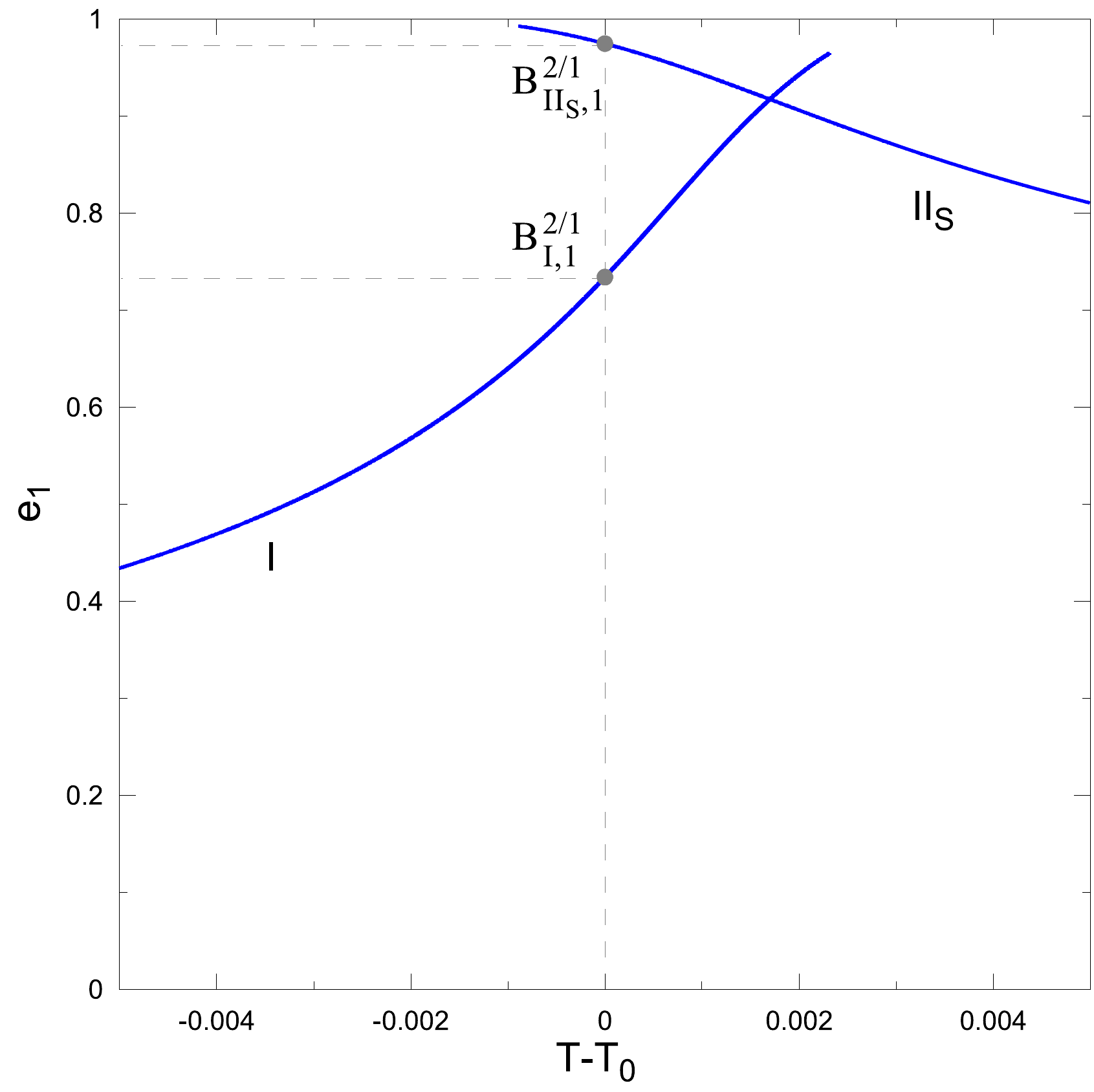}}
\caption{Justification of existence of two bifurcation points, $B^{2/1}_{I,1}$ and $B^{2/1}_{II_S,1}$, in the families ($I$ and $II_S$, respectively) of CRTBP in 2/1 MMR, where $T=T_0=2\pi$, that generate periodic orbits in the ERTBP.}
\label{21bif}
\end{figure}

At points where $T=kT_0$, $k\in\mathbb Z^*$, with $T_0=\frac{2\pi}{|\frac{p+q}{p}-1|}$, along the families of the CRTBP we have a bifurcation point to the ERTBP. We denote them in the following way: $B^{(p+q)/p}_{F,\#}$, in order to distinguish the number of bifurcation points ($\#$) along a certain family ($F$) of a particular MMR ($(p+q)/p$). In Fig. \ref{21bif}, we justify the existence of two bifurcation points, $B^{2/1}_{I,1}$ (known by \citet{vkh09} and called $B^0_T$ therein) and $B^{2/1}_{II_S,1}$ (not reported in previous works). As the period, $T$, varies along the families $I$ and $II_S$ we have bifurcation points when it becomes equal to $T_0=2\pi$. 

\begin{figure}[!h]
\centering
\resizebox{0.65\hsize}{!}{\includegraphics{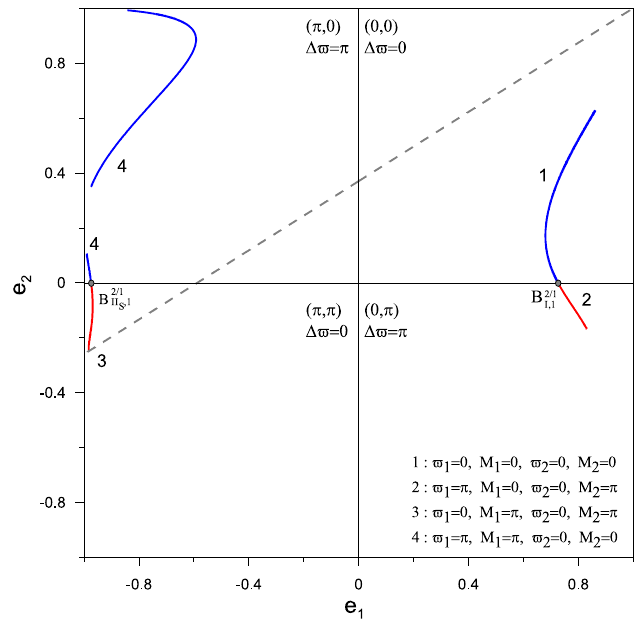}}
\caption{Families of periodic orbits in 2/1 MMR of the ERTBP when $m_2=0.001$ presented as in Fig. \ref{21crtbp} in four quadrants in correspondence with the four different symmetric configurations on the plane $(e_1,e_2)$. Apart from the apsidal difference, $\Delta\varpi$, being noted, the angles in brackets represent the pair of resonant angles ($\theta_1,\theta_2$). The dashed grey curve depicts the collision line between the planets.}
\label{21e_4}
\end{figure}

\begin{figure*}[!ht]
\centering
\includegraphics[height=8cm]{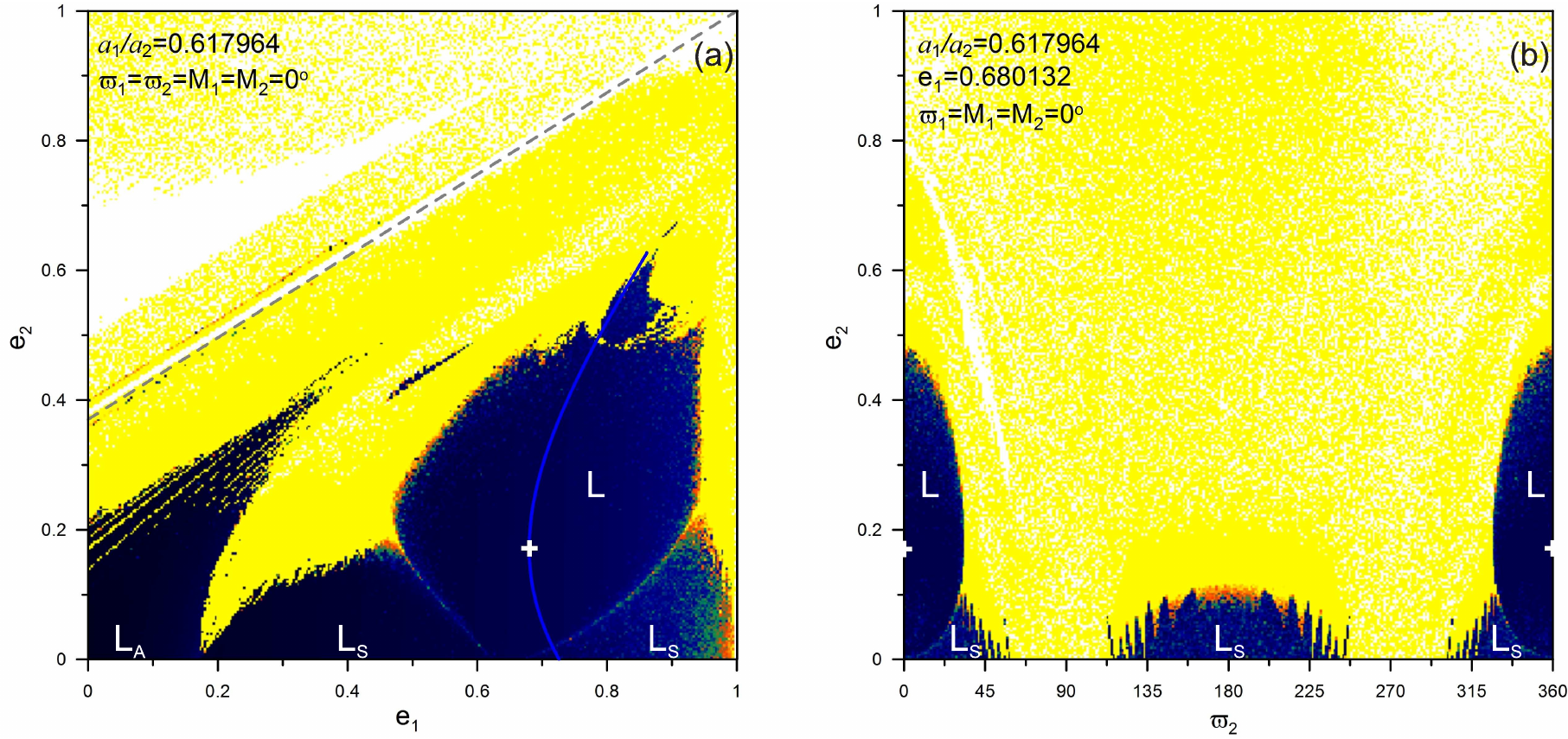}
\caption{DS-maps on the planes \textbf{(a)} ($e_1,e_2$) and \textbf{(b)} ($\varpi_2,e_2$) yielded by a periodic orbit (white cross) that belongs to the stable family (blue curve) of the configuration ($\theta_1,\theta_2$)=($0,0$). The dashed grey line depicts the region where the planetary collisions take place. The orbital elements of the periodic orbit that remain fixed during the computation of each map are noted down. The islands of stability showcase the following attributes: $L$: libration of all resonant angles and apsidal difference (MMR), $L_S$: secondary resonance, $L_A$: apsidal difference oscillation or circulation, and $R$: rotation of both resonant angles and apsidal difference.}
\label{21_000}
\end{figure*}

\begin{figure*}[!ht]
\centering
\includegraphics[height=8cm]{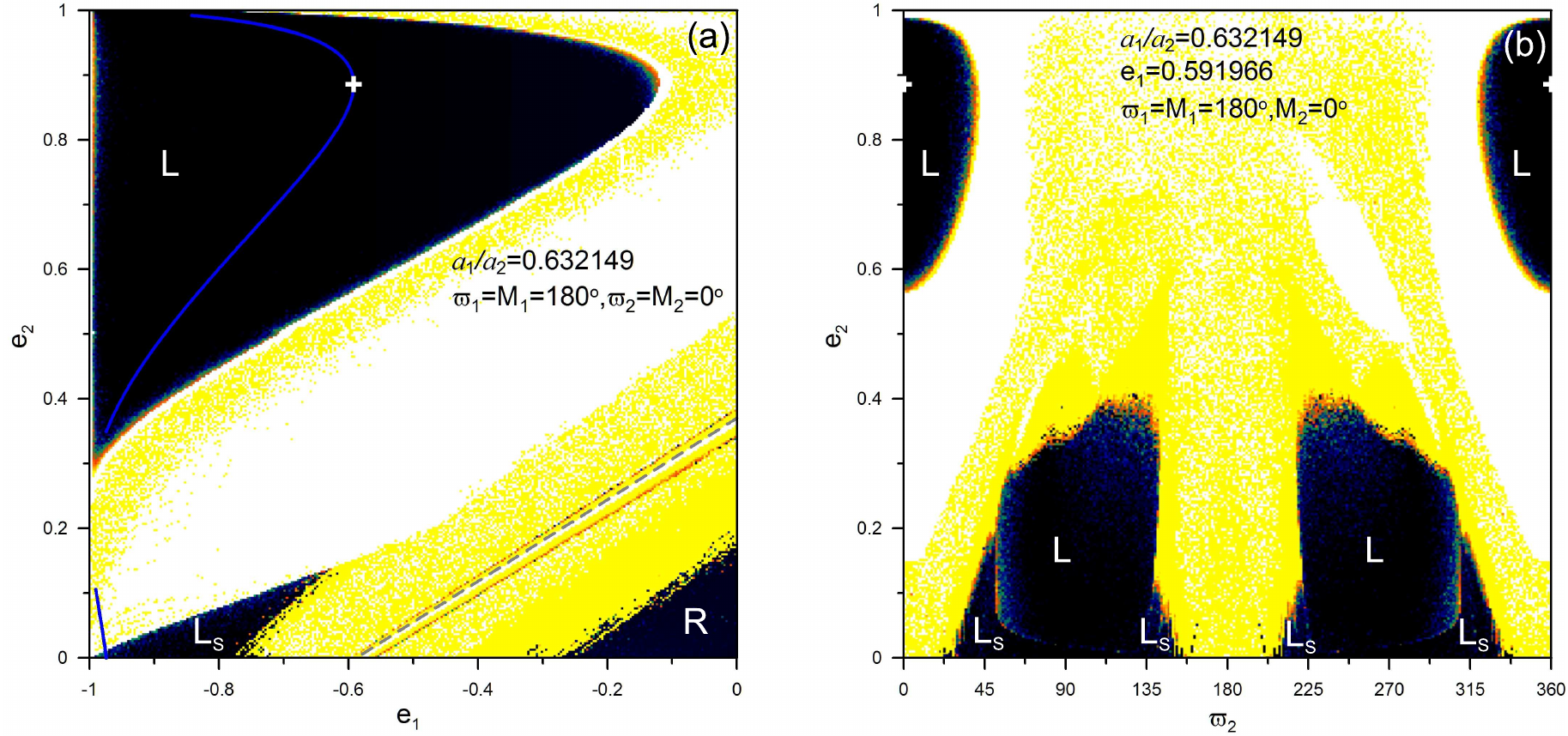}
\caption{DS-maps presented as in Fig. \ref{21_000} guided by a periodic orbit of the configuration ($\theta_1,\theta_2$)=($\pi,0$).}
\label{21_p0p}
\end{figure*}

In the ERTBP, the giant planet is allowed to evolve on an elliptic orbit. Thus, from each bifurcation point two families are generated and correspond to the location of the giant planet at the pericentre and the apocentre. In Fig. \ref{21e_4}, we present the families that are generated from each bifurcation point in four different configurations on the plane $(e_1,e_2)$ by giving a negative value to $e_i$, whenever $\theta_i$, $i=1,2$ librate about $\pi$. The families that emanate from $B^{2/1}_{I,1}$ have already been described by \citet{vkh09} (called $E_{0p}$ and $E_{0a}$ therein). The stable one evolves in the configuration $(0,0)$ while the unstable in ($0,\pi$). From the new bifurcation point $B^{2/1}_{II_S,1}$, one family starts as unstable and evolves in the configuration $(\pi,\pi)$ and the other as stable and belongs to the configuration $(\pi,0)$. The latter family could not be continued as the eccentricity of the inner planet reached very close to one and the integration stopped. We note that this family has stable highly eccentric orbits, even for both planets. 

Given these new families in the ERTBP ($m_1=0$), the origin of the symmetric families for the configuration $(\pi,0)$ in the GTBP for the 2/1 MMR can be completed, if we continue them with respect to the $m_1$ (increase its value). \citet{vkh09} considered both 2/1 and 1/2 MMRs in the ERTBP and showed the origin of the symmetric families for the configuration $(0,0)$ in the GTBP; starting from the 2/1 MMR in ERTBP (as $\rho\rightarrow \infty$), they computed the families for $\rho>1$, which were known by \citet{mbf08a} and starting from the 1/2 MMR in the ERTBP (as $\rho\rightarrow 0$) they computed the families for $\rho<1$ known by \citet{mbf08b}. The symmetric families for the configuration $(\pi,0)$ in the GTBP for the 2/1 MMR were known by \citet{voyhadj05,bmfm06}. Later, \citet{av12} studied these families in both configurations with regard to the bifurcation points for spatial symmetric families in GTBP.

Now, we focus on the regions of stability that appeared in Sect. \ref{maps} for 2/1 MMR and provide a visualisation of the phase space around the stable periodic orbits, about which they are built. By computing DS-maps on the planes $(e_1,e_2)$ and $(\varpi_2,e_2)$, we delineate the boundaries of regular domains.

In Fig. \ref{21_000}, we justify the island of stability that appeared in Fig. \ref{05}a when $e_1=0.5$. Therein, the resonant angles and the apsidal difference librate about 0. Therefore, we delved into the family of that configuration (stable (blue) family of Fig. \ref{21e_4}, when ($\theta_1,\theta_2$)=($0,0$)) and selected the periodic orbit (white cross) with the minimum value of $e_1$ (its orbital elements that remain constant for the computation of each DS-map are given). On the plane $(e_1,e_2)$ in panel a, we observe that the stable (dark coloured) region, denoted by $L$, is built about this stable (blue) family of periodic orbits that provide the position of the MMR. Denoted by $L_S$ are the neighbouring well defined stable regions, where the 1/1 secondary resonance is apparent ($\theta_1$ libration about 0). A regular domain, $L_A$, where the orbits are secured via an apsidal difference oscillation about 0, exists for low values of the eccentricities of both bodies. On the plane $(\varpi_2,e_2)$ in panel b, we observe the main regular domains, $L$ about the stable periodic orbit (the orbit is symmetric at $\Delta\varpi=0$ and thus the island reappears.) Similarly to islands of panel a, beside those (symmetric) domains, the ones exhibiting 1/1 secondary resonance, $L_S$, inside 2/1 MMR appear. For instance, the region $L_S$ centred at $\varpi_2=\pi$ justifies the respective region at Fig. \ref{05}d, where $\theta_1$ librates about 0.
 
In Fig. \ref{21_p0p}, we justify the islands of stability that appeared in Fig. \ref{05}c at highly eccentric and low values of $e_2$. Given the stable (blue) family of the configuration ($\pi,0$) we selected the periodic orbit with the minimum value of $e_1$. In panel a, the very broad region of stability, $L$, is built about it. We further observe an island of 1/1 secondary resonance, where $\theta_1$ librates about $\pi$, linked with the respective one of Fig. \ref{05}c. Furthermore, there exists another domain, $R$, where all of the resonant angles and apsidal difference rotate. In panel b, the domains $L$ centred at $e_2\approx 0.88$ are linked with the configuration of the periodic orbit with $(\theta_1,\theta_2)=(\pi,0)$, $\Delta\varpi=\pi$. The domains, $L$ centred at $\varpi_2=\pi/2$ (or $3\pi/2$ symmetrically) are linked with the configuration $(\theta_1,\theta_2)=(0,0)$, $\Delta\varpi=0$, as such librations take place. Due to the 2/1 MMR such an initial configuration at $t=0$ is equivalent to $(0,0)$ at $t=T/4$. Additionally, due to the multi-dimensionality of the phase space, some sections of the evolution in tori can appear on planes not targeting the configuration showcased. As in Fig. \ref{21_000}b, the 1/1 secondary resonances, $L_S$, are obvious. 

\subsection{Asymmetric periodic orbits}

As shown by \citet{beau94} and \citet{Voyatzis2005a}, asymmetric periodic orbits in the CRTBP exist only in exterior MMRs of the form $1/p, p=2,..$. At these resonances a change of stability is observed along the families of symmetric periodic orbits. At the orbits where this transition is observed we have bifurcation points that generate asymmetric periodic orbits. Since we have focused here on interior MMRs, the 2/1 MMR does not have families of asymmetric periodic orbits in the CRTBP.

\citet{avk11} showed two types of bifurcation points that can generate families of asymmetric periodic orbits in the ERTBP. In the first one, the asymmetric periodic orbits are generated by bifurcation points of the families of symmetric periodic orbits in the ERTBP where the stability changes. In the second, which takes place only at exterior MMRs, asymmetric periodic orbits in the ERTBP can be generated by asymmetric orbits of the CRTBP whose period is a multiple of $2\pi$. For instance, the period at 1/2 MMR should be $4\pi$.

In our study, there is no change of stability along the families of the ERTBP shown in Fig. \ref{21e_4}. Hence, we cannot generate families of asymmetric periodic orbits by bifurcation points of the first type.

However, by searching the phase space, we found 20 initial conditions that correspond to asymmetric periodic orbits at the ERTBP. They are all unstable, exist for highly eccentric orbits of the giant (outer body), and are shown in Table \ref{as} in Appendix~\ref{appendix}.

We have continued these initial conditions and computed the 20 isolated (independent of the families of symmetric periodic orbits) families of asymmetric periodic orbits, which are shown in Figs. \ref{orb1}-\ref{orb4}. Along these families $\Delta\varpi$ and $\Delta M$ are not constant but vary. Hence, in order to visualise the phase space, we not only plot them on ($e_1,e_2$) plane, but also on ($e_1,\Delta M$) and ($e_1,\Delta \varpi$).
The families of asymmetric periodic orbits reported here are unlikely to provide regular domains for the existence of terrestrial planets, in addition to those reported in Sect. 3 for symmetric configurations.

\section{Application to real systems}\label{application}
Any real planetary system, which can be simulated via the ERTBP for interior MMRs, can be located at the DS-maps presented in Figs. \ref{002}-\ref{05} and its long-term orbital stability can be deduced by the periodic orbits existing in their dynamical vicinity, as in Figs. \ref{21_000} and \ref{21_p0p}. Guided by the orbital elements of the celestial bodies and the periodic orbits, the exact boundaries of stable domains can be unravelled. We remind the reader that only (possible) existence of terrestrial planets in symmetric configurations is taken into account here.

\begin{figure*}[!h]
\centering
\includegraphics[width=18cm]{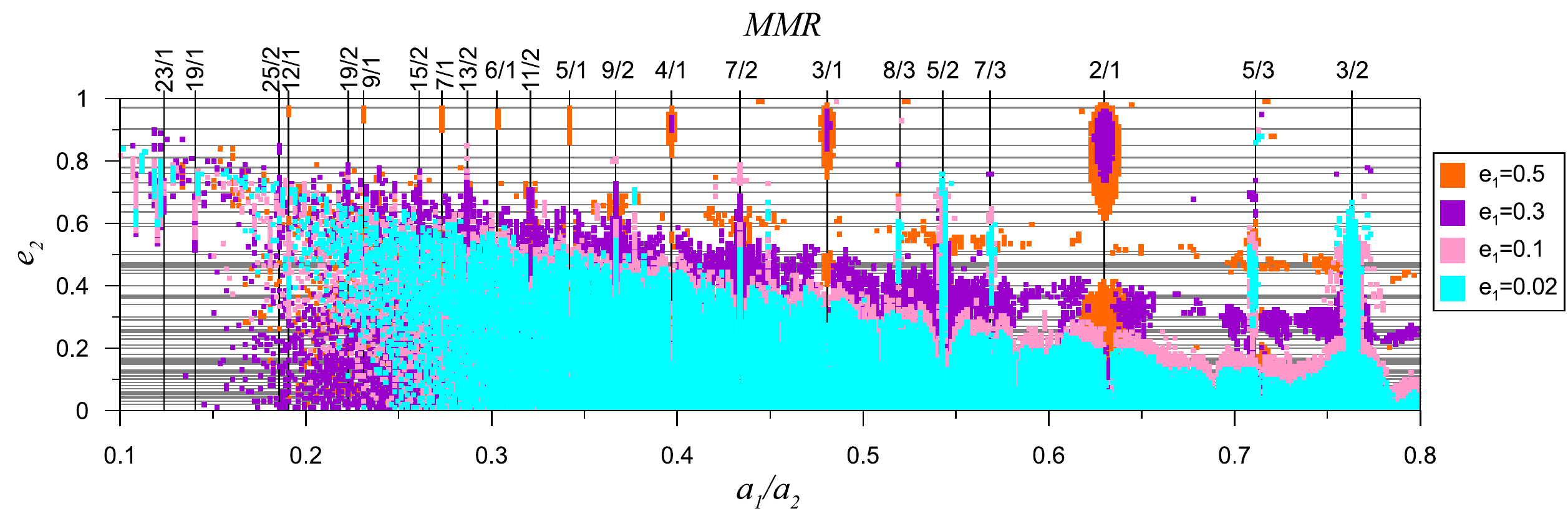}
\caption{All (from Figs. \ref{002}-\ref{05}) long-term stable coplanar symmetric orbits. Emphasis is given to terrestrial planets of low eccentricity values and in case of multiple points on the grid with value $\log(DFLI)\leq 2.5$ only the one with the lowest value $e_1$ is coloured. Plotted in the background with solid grey lines are the 93 planetary systems possessing one giant of mass $[1-5] m_J$ at $a_2\geq 1.0$ AU.}
\label{confall}
\end{figure*}

In Fig. \ref{confall}, we accumulate all long-term stable\footnote{We have broadened the value about which oscillations of the chaotic indicator take place and have set it to $\log(DFLI)\leq2.5$, so that weakly chaotic orbits are included as well. For regular orbits $\log(DFLI)$ should librate about 1.} trajectories originating from the four different symmetric configurations shown in each of the Figs. \ref{002}-\ref{05}. We colour-code the selected orbits from the different figures, and when one point on the plane ($a_1/a_2,e_2$) appears more than once within the selected ones, we keep only the point (colour) that corresponds to the lowest value of $e_1$.
The solid grey horizontal lines represent the 93 planetary systems possessing one detected giant planet of mass within the range $[1-5] m_J$ being located at $a_2\geq 1.0$ AU, as recorded in database {\em{exoplanet.eu}} (October 2017). The solid black vertical lines point out the main MMRs. 

\subsection{For a telescope proposal}\label{sub51}
With telescopes having the instrumental precision needed at hand, the observational astronomers might like to find an Earth-like planet in a system already hosting a giant. The grey lines in Fig. \ref{confall} correspond to the discoveries to-date and hence, new lines can be added, as new ones are brought to light. Observers might need to make a choice between those systems and prioritise them within a given time for a telescope proposal. Based on the eccentricity of the giant planet, $e_2$, and its errors, two strategies are required: 
\begin{enumerate}[I]
\item If $e_2 \geq 0.4$, the regions of stability are small yet distinct. Thus, by looking at the specific $a_1$ of those domains, the orbital elements of the terrestrial planet -if it exists and is found- could be well constrained through dynamical analyses. In particular, terrestrial planets can only be found in eccentric orbits ($e_1\geq 0.3$) and in MMRs (mainly 2/1, 3/1, 4/1, and 5/1). 
\item If $e_2 < 0.4$, the region of stability is very broad and encompasses both non-resonant and resonant motion. Therefore, on the one hand, the possibility of finding a terrestrial planet here might be higher. On the other hand, it becomes harder, as $a_1$ can span a large domain. 
\end{enumerate}
Next, we elaborate on the minimum value of $e_1$ and the domains that can host a terrestrial planet for the single-giant planet systems observed so far.

\subsection{For the single-giant planet systems to-date}\label{sub52}

We divide the 93 detected systems discussed here into the above categories, namely the systems whose giant planet has eccentricity values $e_2\geq 0.4$ (22 systems) and all the rest. For detected giant planets with low to moderate eccentricity, namely $e_2 < 0.4$, the possible locations of a terrestrial planet in the planetary system are numerous, as the grey lines cross the broad main region of stability. If not locked in an MMR and thus, be located within a tongue, secondary resonances and apsidal resonance provide the long-term stability of a possible terrestrial planet. Non-resonant orbits that belong to domain $A$, below the collision line, will not intersect. Therefore, we observe that this area is largely populated by regular orbits where $e_1=0.02$ or 0.1. We note again that at those eccentricities we can have no trapping in 5/1, 4/1, 3/1, and 2/1 MMRs, but only in 5/2 and 3/2 MMRs (see also the description in Sect. \ref{maps} and Figs. \ref{002} and \ref{01}). Regarding the 22 systems  for which $e_2\geq 0.4$, we report in Table \ref{tab} (appendix \ref{possible}), the possible MMRs and values of the eccentricities allowing the long-term existence of a terrestrial planet. There is a tendency of possible survival in 3/2, 2/1, 5/2, 3/1, and 4/1 MMRs rather than in higher order MMRs, because either the regions of stability are narrower or the giant is located at the borders.

To begin with, in HD 20782 \citep{20782} with $e_2=0.97\pm 0.01$ a terrestrial planet could survive if locked in 2/1 MMR evolving in an elliptic orbit $e_1\geq 0.3$. Trappings in 3/1, 4/1, 5/1, 6/1, 7/1, 9/1, and 12/1 MMRs are possible, but survival is not probable, since the giant is located at the borders of those islands of stability - far away the periodic orbit. Hence, the oscillations of the orbital elements would be of very large amplitude. The same holds for the system HD 108341 \citep{108341} with $e_2=0.85\pm 0.09$.
In HD 4113 \citep{4113} with $e_2=0.903\pm 0.005$ a terrestrial planet could survive if locked in 2/1, 3/1, and 4/1 MMRs evolving in an elliptic orbit $e_1\geq 0.3$. 2/1 MMR is again favoured, since the island is broader, but also 3/1 and 4/1 captures are likely to support the survival.

The systems HD 28254 \citep{28254} with $e_2=0.81\pm 0.02$, HD 45350 \citep{45350} with $e_2=0.778\pm 0.009$ and  HD 30562 \citep{30562} with $e_2=0.76\pm 0.05$ could dynamically host a terrestrial planet with $e_1\geq 0.3$ at 2/1 and 3/1 MMRs. For HD 28254 there could be a capture at 9/2 ($e_1=0.02$) and 13/2 ($e_1\geq 0.02$) MMRs and given the errors at 7/2 ($e_1=0.02$), 4/1 ($e_1=0.5$) and 25/2 ($e_1=0.3$) MMRs. As for HD 45350 a possible capture could be achieved at 8/3 ($e_1=0.3$), 7/2 ($e_1=0.02$), 15/2 ($e_1=0.3$) and 19/2 ($e_1=0.3$) MMRs. Additionally, given the error of the eccentricity of HD 30562, a terrestrial planet could be apparent at 5/2 MMR for $e_1=0.02$ and other MMRs, like 19/2, 15/2, 13/2, and 7/2 with $e_1=0.1$ and 23/1, 25/2, 19/2, 15/2, 11/2, 9/2, and 7/3 with $e_1=0.3$, but its long term survival would be dependent on precise orbital elements (close to the respective periodic orbits), as those stability domains (tongues) are very narrow.

Likewise HD 30562, the systems HD 86226 \citep{86226129445} with $e_2=0.73\pm 0.21$, HD 129445 \citep{86226129445} with $e_2=0.7\pm 0.1$ could host a terrestrial planet trapped in all of the above-mentioned MMRs; most probable being the 2/1 ($e_1=0.5$) and 5/2 ($e_1\geq 0.02$) MMRs. Nonetheless, the error at $e_2$ in HD 86226 is very large and based on that range, another planet could also be dynamically locked in its vicinity at 25/2 MMR with $e_1=0.1$, at 8/3, 7/3, 5/3, 3/2 MMRs with $e_1\geq 0.02$ at 3/1, 7/2, 9/1, 11/2, 13/2, 15/2 MMRs with $e_1=0.3$ and at 4/1, 5/1, 6/1, 7/1, 9/1, 12/1 MMRs with $e_1=0.5$. The error at HD 129445 could allow captures in the above MMRs as well (apart from 3/1, 4/1, and 6/1) but also in 12/1 (for $e_1=0.1$) and 25/2 (for $e_1=0.3$) MMRs.
 
The giant planets of HD 120084 \citep{120084} with $e_2=0.66\pm 0.1$, HD 16175 \citep{16175} with $e_2=0.637\pm 0.02$ and HD 152079 \citep{86226129445} with $e_2=0.6\pm 0.24$ could be in 2/1 MMR with a terrestrial planet only if $e_1\geq 0.3$. By considering the errors about the central values of the eccentricities, a capture can occur at 3/1 MMR when $e_1=0.5$, but also at the islands and tongues at 19/1, 19/2, 15/2, 13/2, 11/2, 9/2, and 7/2 MMRs, where $e_1=0.3$, at 25/2 MMR where $e_1=0.1$ and 8/3, 5/2, 7/3, 5/3, and 3/2 MMRs, where $e_1\geq 0.02$. The broader regular domains (and therefore possible existence and survival) are centred at 5/2, 2/1, and 3/2 MMRs. The error of HD 152079 could allow captures in 3/1 MMR for $e_1=0.3$, 4/1 and 5/1 MMRs for $e_1=0.5$. The above captures hold also for the systems HD 171028 \citep{171028} with $e_2=0.59\pm 0.01$, HD 79498 \citep{79498} with $e_2=0.59\pm 0.02$ with the only exception that these two systems could exhibit a capture in 3/1, 4/1, and 5/1 MMRs.
 
The planets of HD 220773 \citep{79498} with $e_2=0.51\pm 0.1$, HD 142415 \citep{142415} with $e_2=0.5$, HD 29021 \citep{29021} with $e_2=0.459\pm 0.008$, HD 210277 \citep{210277} with $e_2=0.472\pm 0.011$, HD 66428 \citep{66428} with $e_2=0.465\pm 0.03$, HD 213240 \citep{213240} with $e_2=0.45\pm 0.04$, HD 23127 \citep{23127}) with $e_2=0.44\pm 0.07$, HD 162004 \citep{162004} with $e_2=0.4\pm 0.05$ and HD 171238 \citep{171238} with $e_2=0.4\pm 0.065$ could also host in their dynamical neighbourhood an exo-Earth with $e_1=0.02$ or terrestrial planets with greater eccentricity values. We refer the reader to Figs. \ref{002}-\ref{05} to determine the precise range of each tongue mentioned herein.

\section{Discussion and conclusions}
\label{con}

In this work, we have studied the dynamical stability of a giant outer planet and an inner terrestrial one and model their evolution with the ERTBP. We conclude the long-term stability and thus, possible existence and survival of such a coexistence by computing a chaotic indicator, the DFLI. We realised a broad exploration of phase space, indicate the domains where Keplerian ellipses intersect and justify the stable regions via the stable periodic orbits by using as an example the 2/1 MMR.  

In particular, there are three well-known types of resonance that can guarantee a regular evolution and are observed on the DS-maps of Fig. \ref{confall}:
\begin{enumerate}
				\item the MMRs (or ACRs), where both the resonant angles and the apsidal difference librate about specific values determined by the periodic orbits, which buttress regular domains. Those regions form tongues or distinct islands at low to moderate or highly eccentric values of $e_2$, respectively. 
				\item the secondary resonances, where only $\theta_1$ librates, while $\theta_2$ and $\Delta\varpi$ rotate.
				\item the apsidal resonance, where only $\Delta\varpi$ oscillates, while the rest resonant angles rotate.
\end{enumerate}
In all of the above, close encounters (even when the orbits do intersect) are avoided as the phases are protected.

The families of periodic orbits constitute the backbone of stability domains in phase space and they can act as a diagnostic tool, which can help ascertain information with regards to the dynamical vicinity of an exoplanet. As shown here, they can help predict a possible survival of a celestial body in the dynamical neighbourhood of already known exoplanets. For the 2/1 MMR in the CRTBP we found a new bifurcation point which in turn allowed us to compute two new families of symmetric periodic orbits in ERTBP. We also computed 20 new isolated families of asymmetric periodic orbits in the ERTBP. Although the initial conditions of the asymmetric periodic orbits are unstable, the unstable manifolds can be applied for instance to trajectory design \citep[see e.g.][]{broad}. We herein focused on 2/1 MMR, but a respective study for the rest major MMRs will follow. 

We pursued a study of all coplanar symmetric configurations for semi-major axes ratios within the range [0.1,0.8], so that we study interior MMRs apart from the non-resonant cases. We considered Earth-like planets and moderately eccentric ones with $e_1=0.02$ and 0.1, 0.3, and 0.5, respectively. 

After having gathered the (93) giant planets detected so far with mass [1-5]$m_J$ with no terrestrial companions to-date and located at $a_2\geq 1.0$ AU, we can draw the following conclusions:
\begin{enumerate}
        \item When the outer giant planet is highly eccentric, a terrestrial planet can only survive if locked in MMR with it (distinct islands). Its orbit has to be quite eccentric. For instance, in 2/1 MMR it has to be $>0.2$.
        \item In the rest of the resonant cases when the giant has lower eccentricity values, the inner terrestrial planet can survive when being either in MMR (within the tongues for circular up to moderately eccentric orbits of the giant), or in secondary resonance (for $e_2<0.2$).
        \item The broad stable region below the collision line (domain $A$) also consists of non-resonant regular orbits, where the stability is guaranteed. We note that there exist weakly chaotic orbits for low semi-major axes ratios, not depicted at Fig. \ref{confall} (see the blue-green region in Figs. \ref{002}-\ref{05}), where a terrestrial planet may survive long-time spans. 
\end{enumerate}

For each planetary system of the above, where additionally $e_2>0.4$, i.e. for 22 of them, we provide some possible resonant solutions. Our results can be applied to any system that can be modelled by the ERTBP, namely star-asteroid-giant planet, planet-satellite-spacecraft, binary stars-circumprimary (S-type) planet or like herein star-terrestrial planet-giant planet. In particular, based on the MMRs, observations can be driven to specific semi-major axes ratios, so that a possible existence of a terrestrial planet could be revealed (Sect. \ref{sub51}). Our study is also helpful for detected systems consisting of an inner terrestrial planet and an outer giant planet (Sect. \ref{sub52}), to restrict or complement the observational data. \\

\noindent
{\bf Acknowledgements.} \\ The work of KIA was supported by the Fonds de la Recherche Scientifique-FNRS
under Grant No. T.0029.13 (“ExtraOrDynHa” research project). Computational resources have been provided by the Consortium des Équipements de Calcul Intensif (CÉCI), funded by the Fonds de la
Recherche Scientifique de Belgique (F.R.S.-FNRS) under Grant No.2.5020.11. 

\bibliographystyle{aa} 

\bibliography{nbibab} 
\clearpage
\begin{appendix} \setcounter{figure}{0} \renewcommand{\thefigure}{A.\arabic{figure}} 
\section{Families of asymmetric periodic orbits in 2/1 MMR in the ERTBP}

\label{appendix}

\begin{figure}[!h]
\centering
$\begin{array}{c}
 \rm{Family \; 1}  \\
\includegraphics[height=2.9cm]{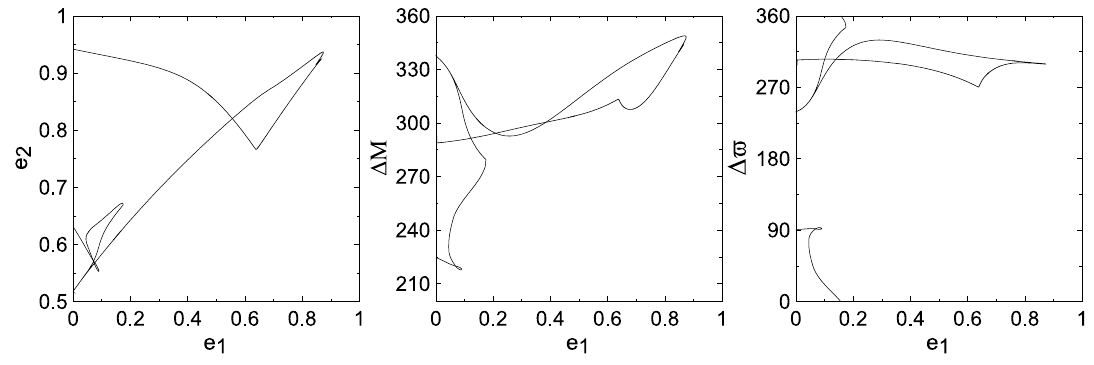}\\
 \rm{Family \; 2}  \\
\includegraphics[height=2.9cm]{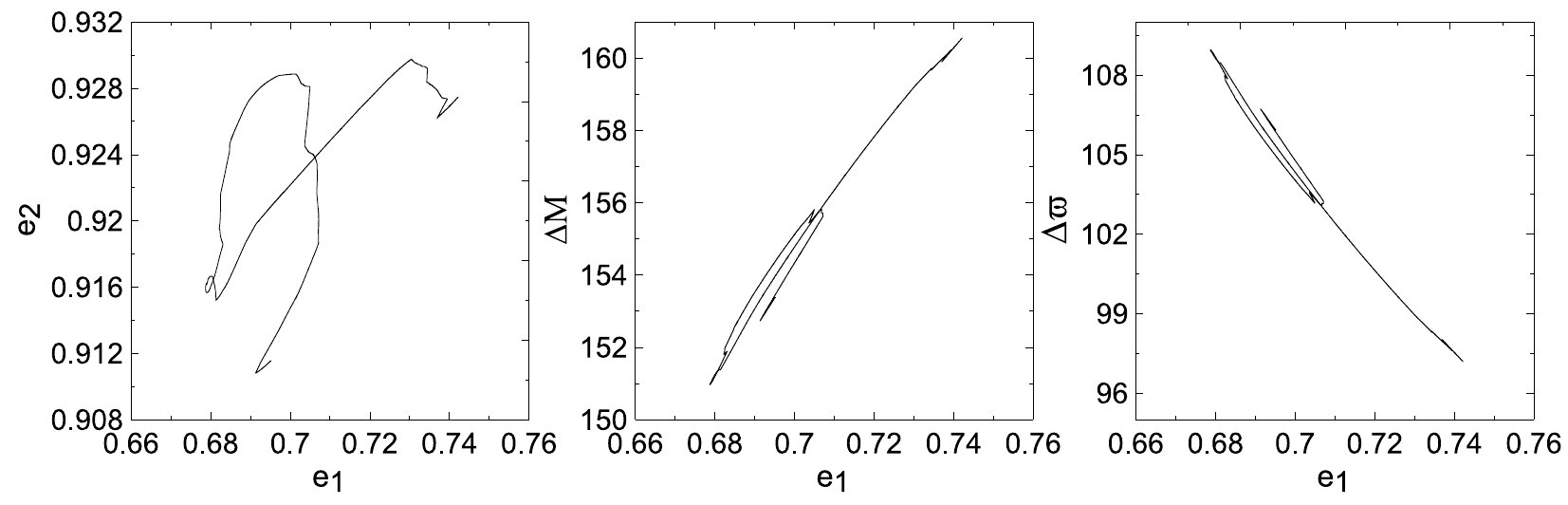}\\
 \rm{Family \; 3}  \\
\includegraphics[height=2.9cm]{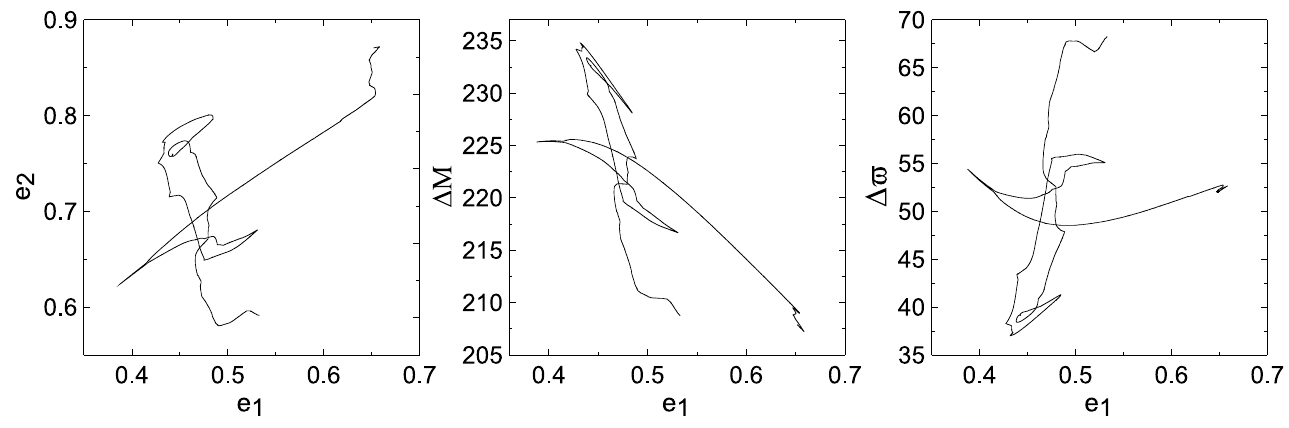}\\
 \rm{Family \; 4}  \\
\includegraphics[height=2.9cm]{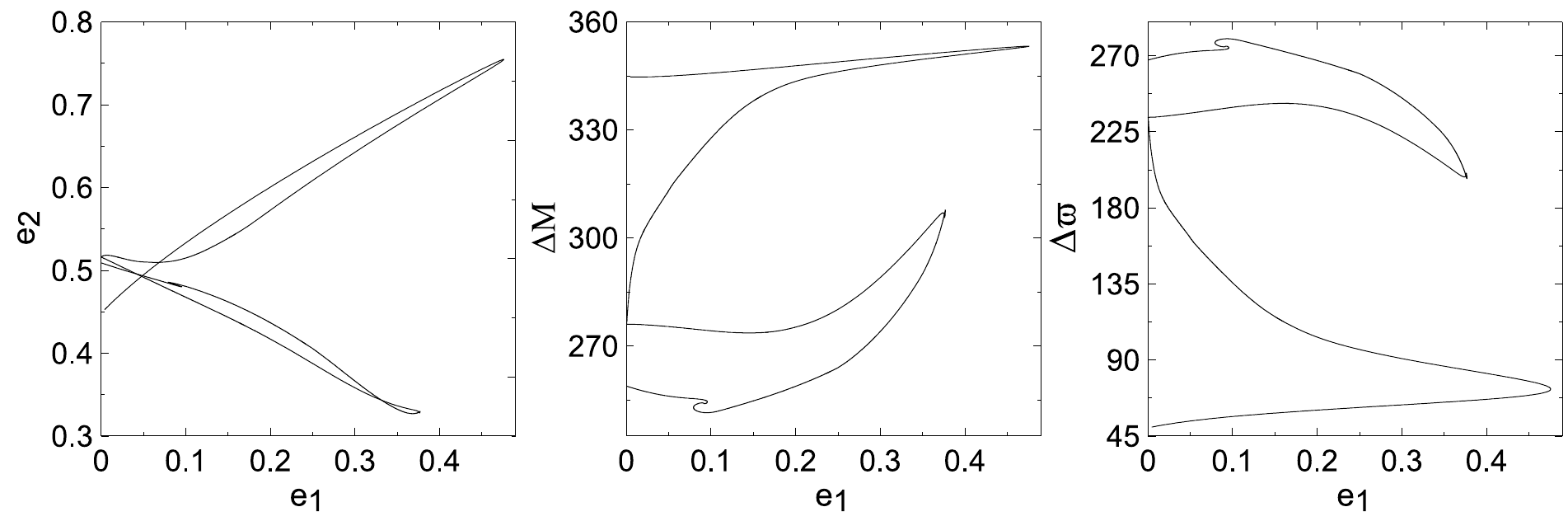}\\
 \rm{Family \; 5}  \\
\includegraphics[height=2.9cm]{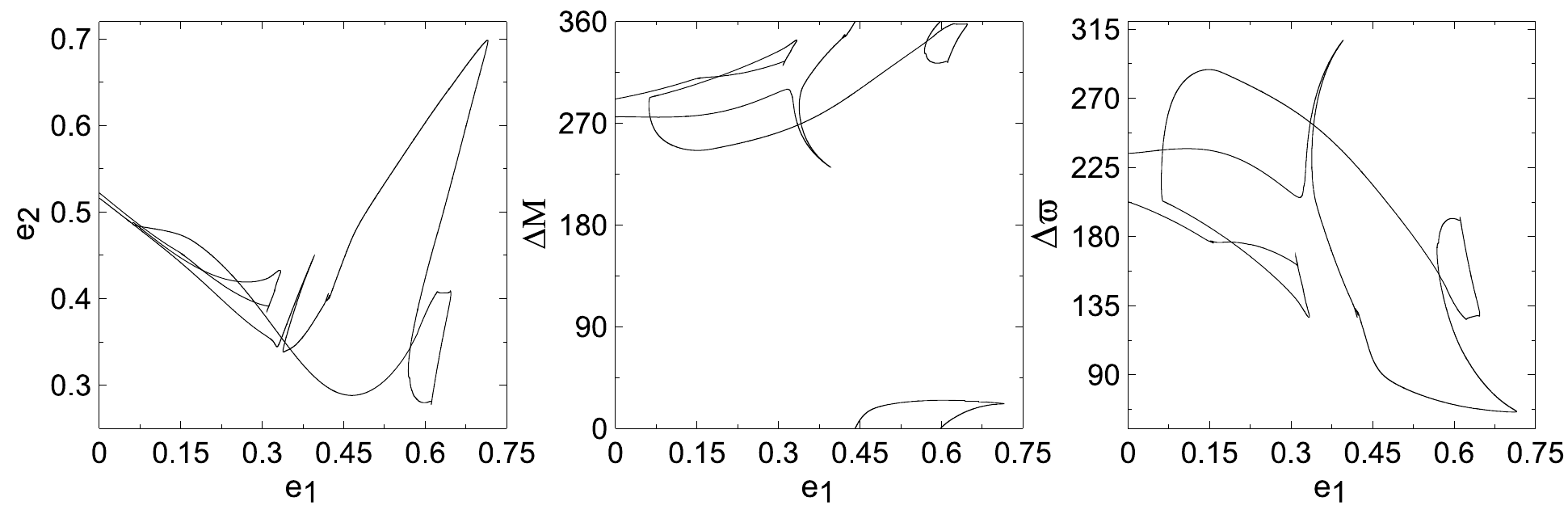}\\
\end{array} $
\caption{Isolated families of asymmetric periodic orbits in ERTBP for the 2/1 MMR presented on the following planes $(e_1,e_2)$, $(e_1,\Delta M)$ (dynamic asymmetry) and $(e_1,\Delta\varpi)$ (geometric asymmetry). The initial conditions are given in Table \ref{as} and are all unstable. These families are independent from the families of symmetric periodic orbits in the ERTBP.}
\label{orb1}
\end{figure}

\begin{figure}[!h]
\centering
$\begin{array}{ccc}
 \rm{Family \; 6}  \\
\includegraphics[height=2.9cm]{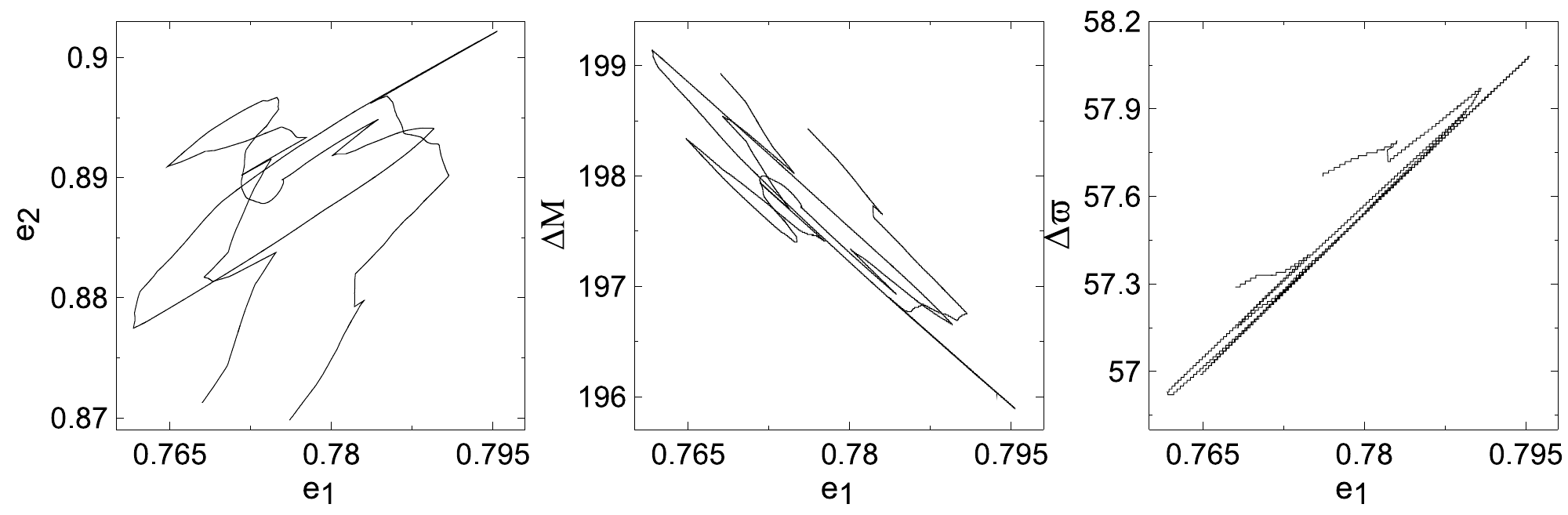}\\
 \rm{Family \; 7}  \\
\includegraphics[height=2.9cm]{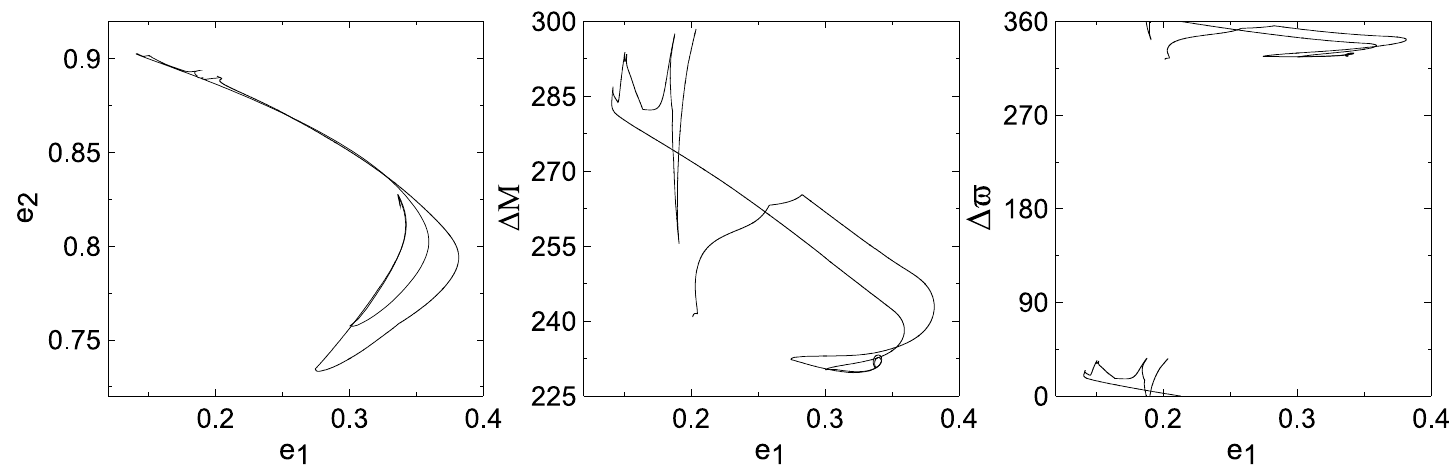}\\
 \rm{Family \; 8}  \\
\includegraphics[height=2.9cm]{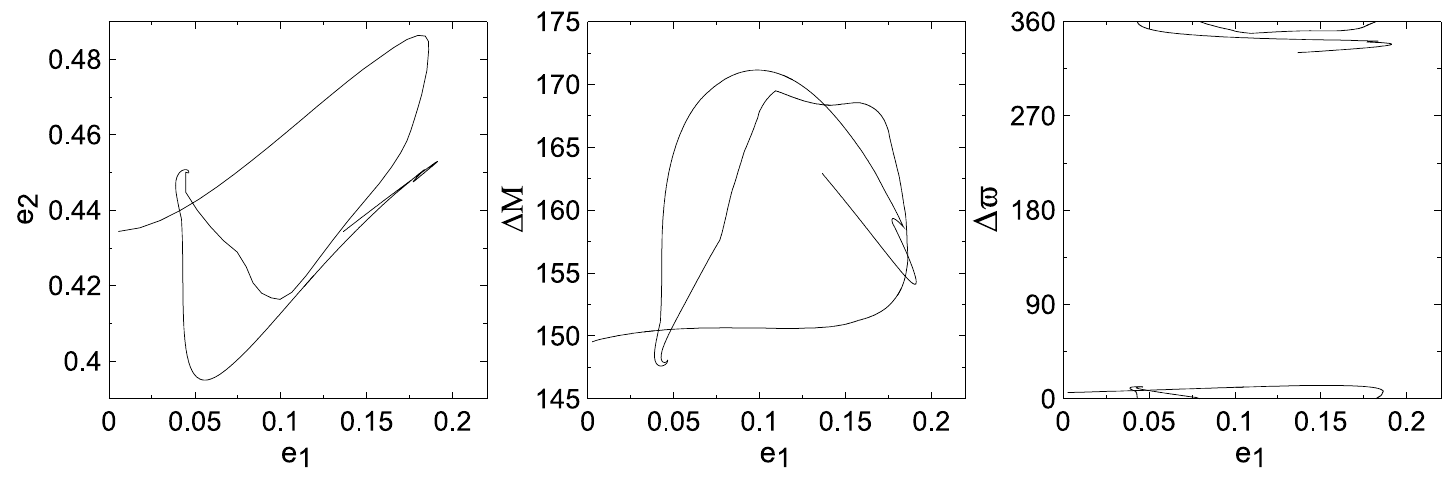}\\
 \rm{Family \; 9}  \\
\includegraphics[height=2.9cm]{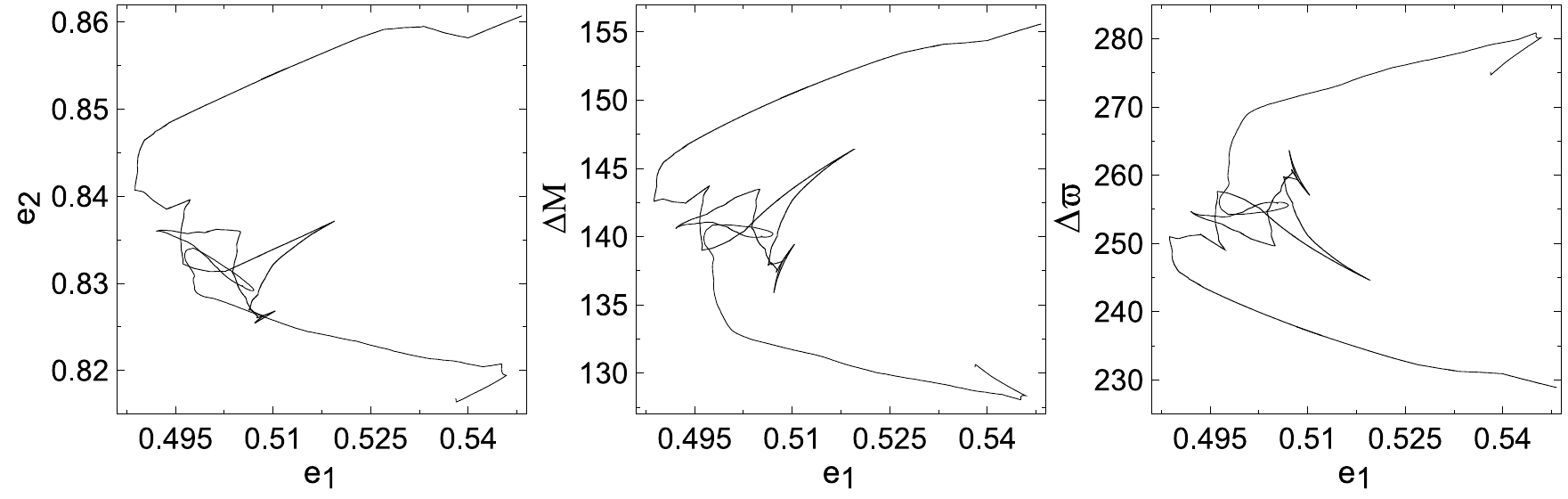}\\
 \rm{Family \; 10}  \\
\includegraphics[height=2.9cm]{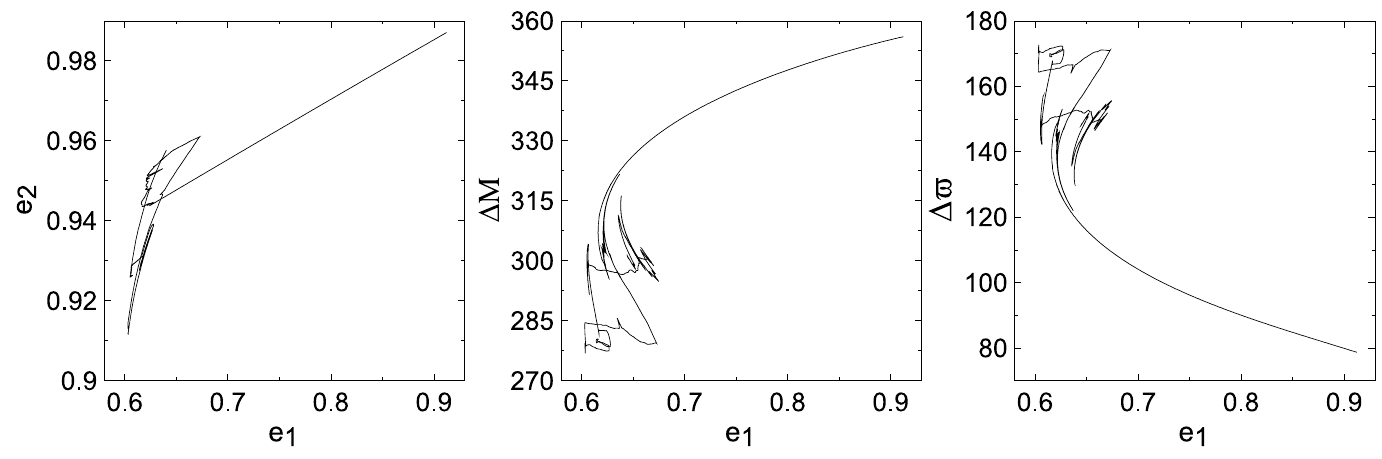}\\

\end{array} $
\caption{Families 6-10, as in Fig. \ref{orb1}.}
\label{orb2}
\end{figure}

\begin{figure}[!h]
\centering
$\begin{array}{c}
 \rm{Family \; 11}  \\
\includegraphics[height=2.9cm]{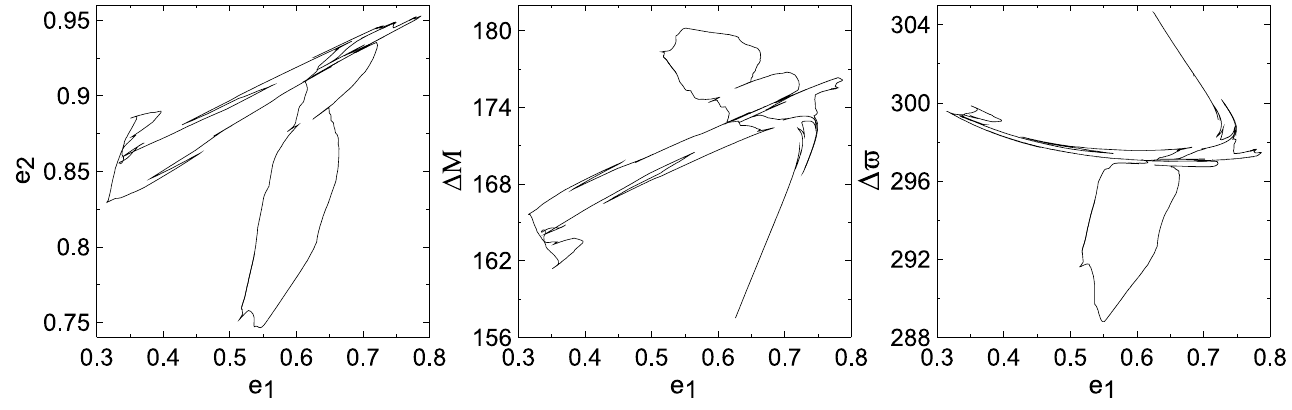}\\
 \rm{Family \; 12}  \\
\includegraphics[height=2.9cm]{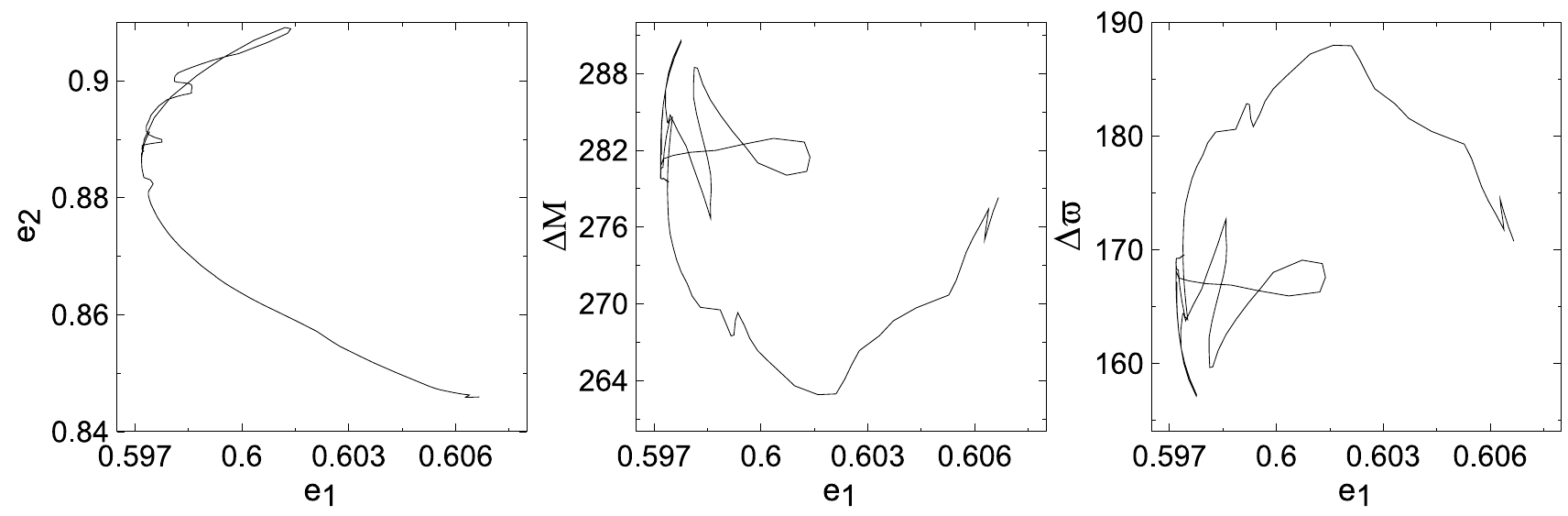}\\
 \rm{Family \; 13}  \\
\includegraphics[height=2.9cm]{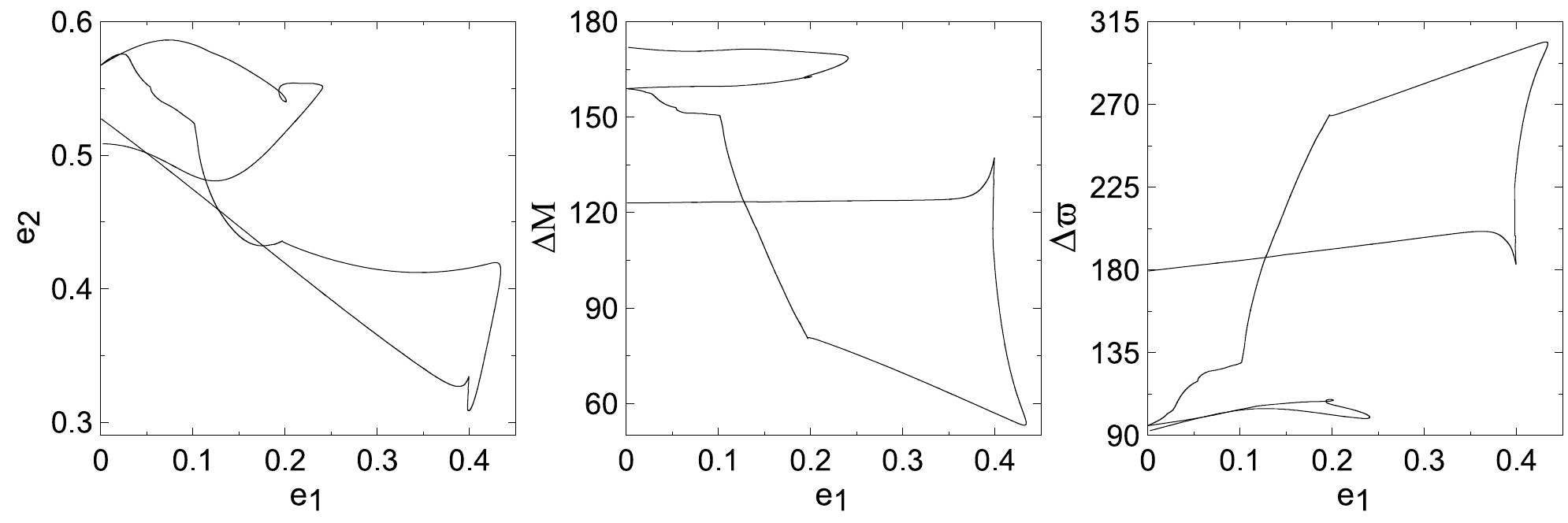}\\
 \rm{Family \; 14}  \\
\includegraphics[height=2.9cm]{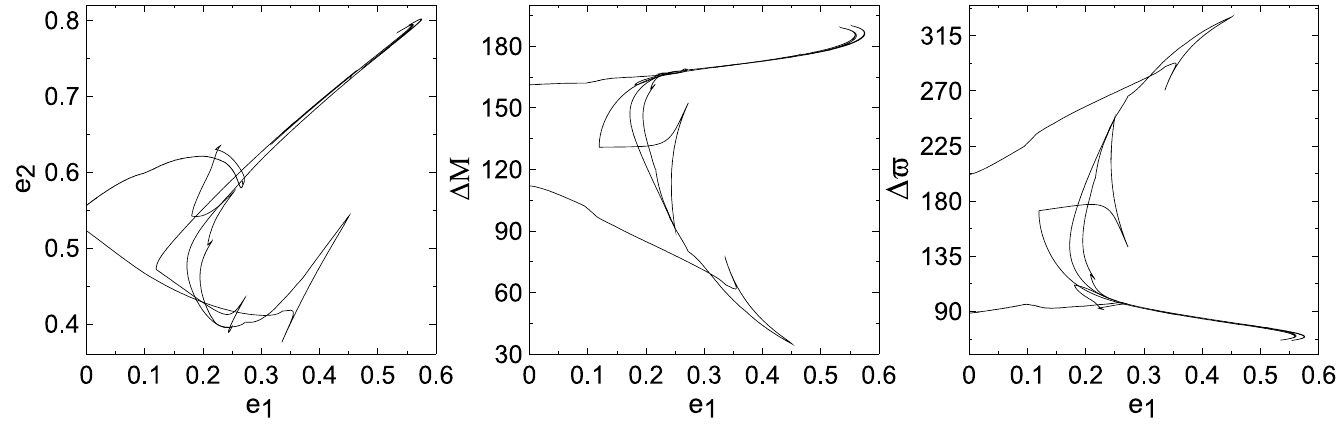}\\
 \rm{Family \; 15}  \\
\includegraphics[height=2.9cm]{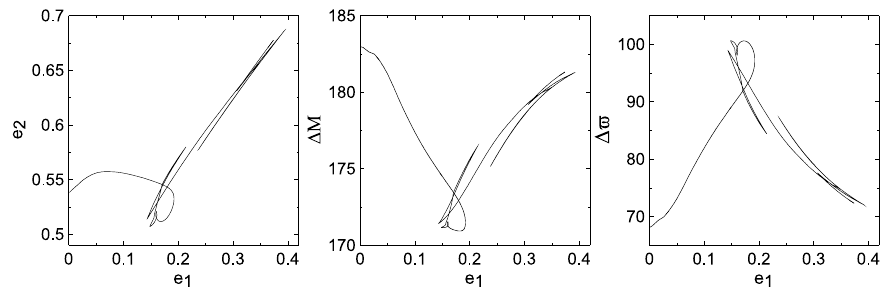}\\

\end{array} $
\caption{Families 11-15, as in Fig. \ref{orb1}.}
\label{orb3}
\end{figure}

\begin{figure}[!h]
\centering
$\begin{array}{c}
 \rm{Family \; 16}  \\
\includegraphics[height=2.9cm]{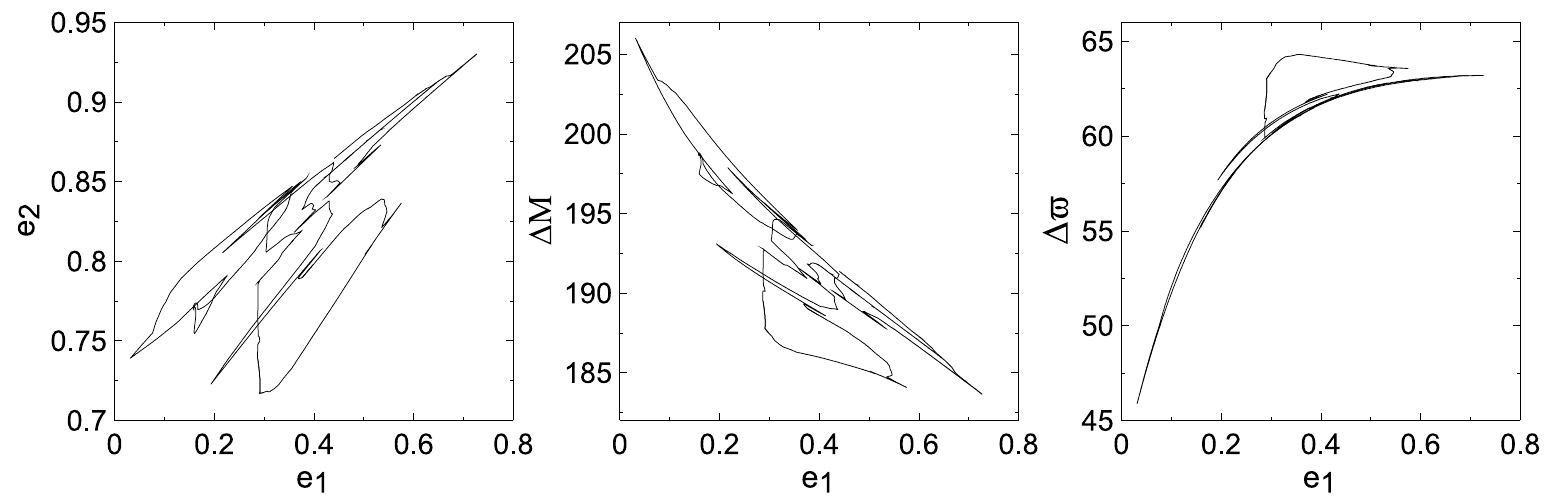}\\
 \rm{Family \; 17}  \\
\includegraphics[height=2.9cm]{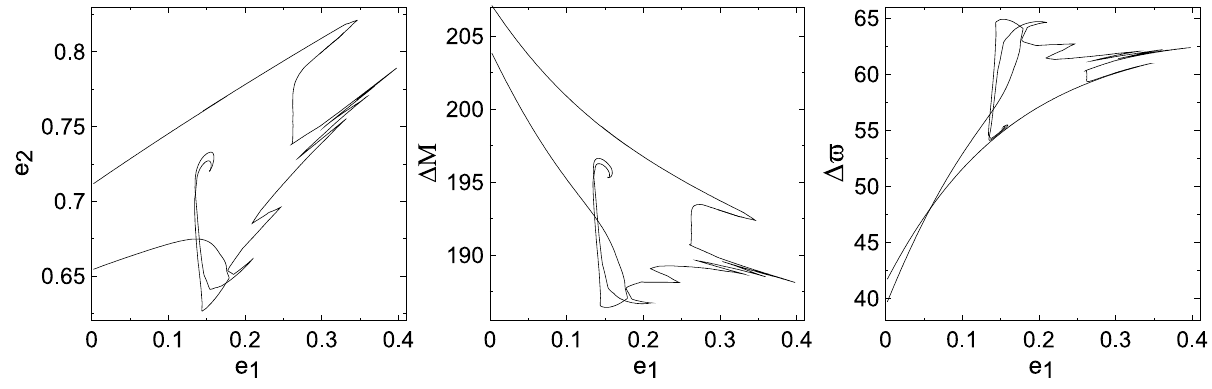}\\
 \rm{Family \; 18}  \\
\includegraphics[height=2.9cm]{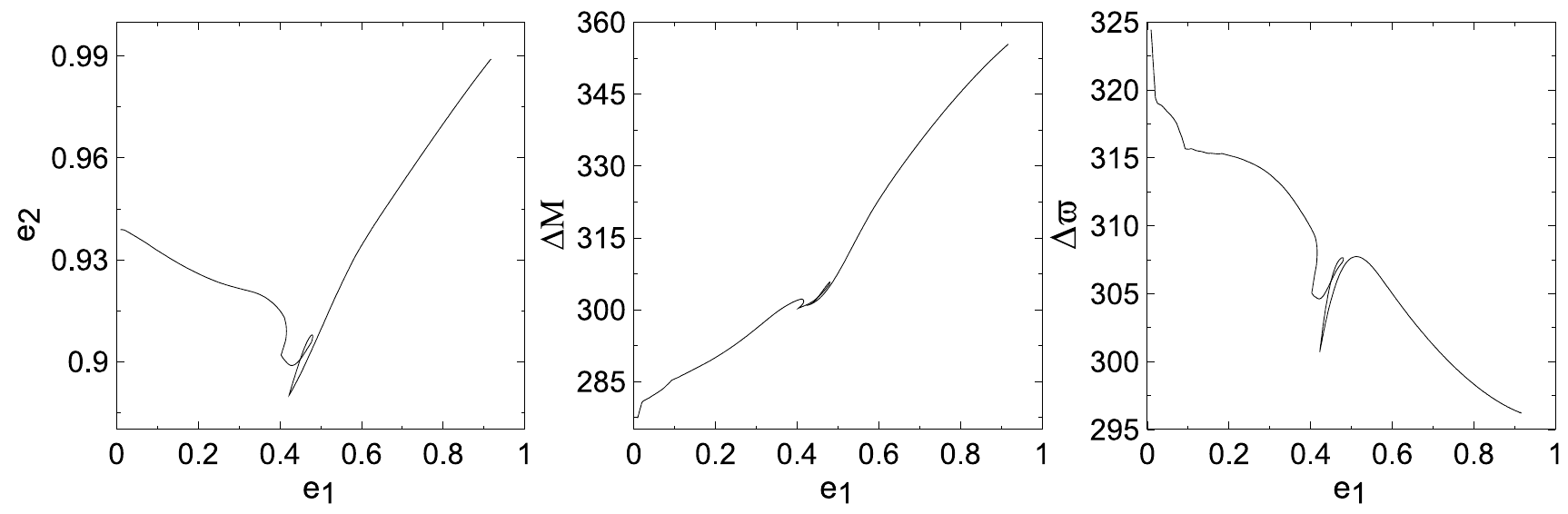}\\
 \rm{Family \; 19}  \\
\includegraphics[height=2.9cm]{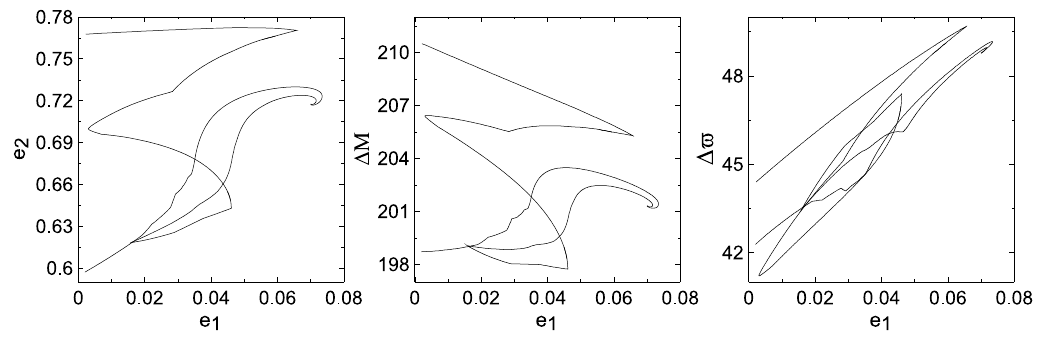}\\
 \rm{Family \; 20}  \\
\includegraphics[height=2.9cm]{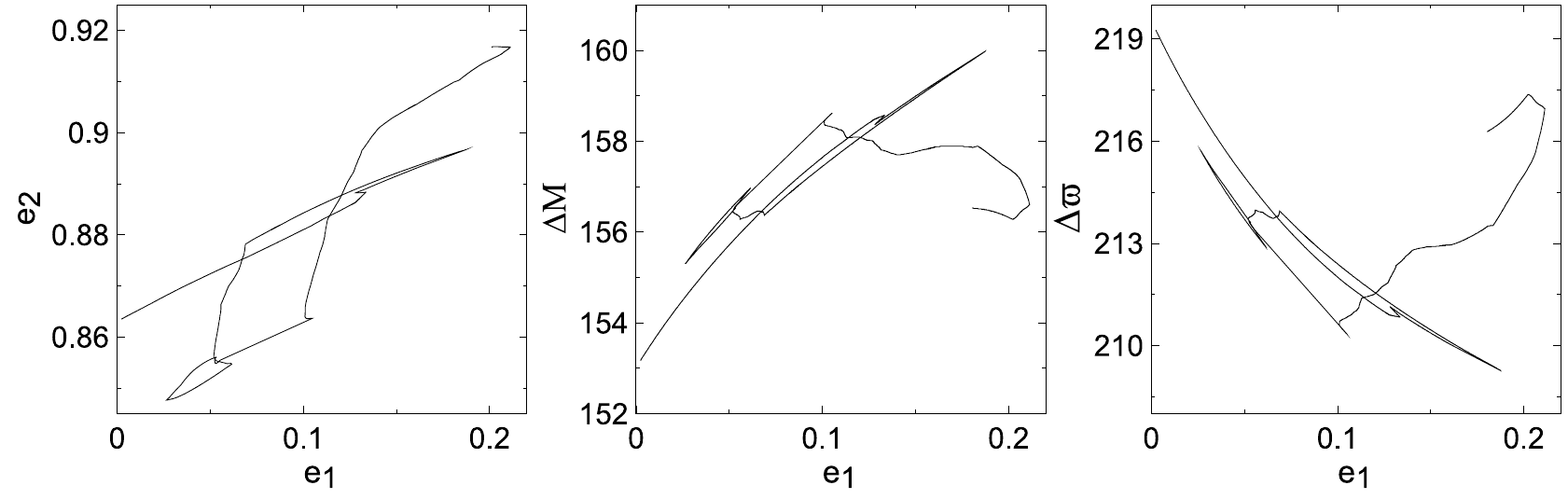}\\

\end{array} $
\caption{Families 16-20, as in Fig. \ref{orb1}.}
\label{orb4}
\end{figure}
\clearpage

\begin{table*}[!htb]
\caption{Initial conditions of the families 1-20 of asymmetric periodic orbits in 2/1 MMR.}
\scalebox{0.58}{\begin{tabular}[b]{ccccccccccc}
\toprule
Family&$x'$&$x$&$y$&$\dot x$&$\dot y$&$\dot\theta$&$e_1$&$e_2$&$\Delta M$ ($^{\circ}$)&$\Delta\varpi$ ($^{\circ}$) \\
 \cmidrule{1-11}
1 & 0.72331794750700 & 0.49013195157962 & -1.82695158386230 & -2.65478370333012 & -0.56701605244671 & 1.69319527200545 & 0.0860163 & 0.5580729 & 218.65 & 91.50 \\
 \cmidrule{2-11}
2 & 0.63321025768200 & 1.59453165011230 & -5.67590694541681 & -14.45329652263383 & -4.13526567777250 & 2.57025981145974 & 0.6789292 & 0.9161948 & 151.02 & 108.89 \\
 \cmidrule{2-11}
3 & 0.63047186508700 & -1.88248865529942 & -2.29632923979982 & -5.16928490819296 & 4.46273668460205 & 2.39803573831806 & 0.4425901 & 0.7593570 & 232.63 & 39.13 \\
 \cmidrule{2-11}
4 & 0.63087376727600 & -0.87705151805361 & 1.12691498748010 & 2.01174310935965 & 1.21087189099085 & 2.08389129602805 & 0.0914706 & 0.4816108 & 254.06 & 275.54 \\
 \cmidrule{2-11}
5 & 0.63420301291300 & 0.29197254892169 & 1.00974268447896 & 1.13065744890431 & -0.79803090077487  & 2.12456885857444 & 0.1525533 & 0.4513824 & 309.80 & 176.68 \\
 \cmidrule{2-11}
6 & 0.64118423931800 & -4.29365702951665 & -7.33451899022320 & -18.90829479679702 & 11.11113087802684  & 2.59393188047815 & 0.7754161 & 0.8898140 & 197.72 & 57.38 \\
 \cmidrule{2-11}
7 & 0.71748622583200 & -2.70800622071297 & 1.20843227804079 & -2.24367758267459 & -5.45250914881905  & -1.90233251403904 & 0.3379752 & 0.8213934 & 230.83 & 326.88 \\
 \cmidrule{2-11}
8 & 0.63482968402900 & -1.49288772007308 & 0.06759364911299 & -0.00782516489640 & 2.40969738972048  & 2.02132884454501 & 0.0463576 & 0.4498677 & 148.07 & 11.19 \\
 \cmidrule{2-11}
9 & 0.65735722962700 & 1.33246033480411 & 3.48610081445896 & 7.86138858484334 & -2.87474461680602  & 2.30083405955404 & 0.5052528 & 0.8297413 & 140.02 & 255.90 \\
 \cmidrule{2-11}
10 & 0.63399837094100 & 3.98254236120459 & 0.05416001257745 & -0.17928506911680 & -9.87884493186210  & 2.51705575134906 & 0.6161609 & 0.9297257 & 280.26 & 169.18 \\
 \cmidrule{2-11}
11 & 0.63283580598800 & -1.26763539799044 & 2.49305681758722 & 5.49108002978163 & 2.85458239643100  & 2.29401432064642 & 0.3350573 & 0.8560322 & 164.23 & 299.33 \\
 \cmidrule{2-11}
12 & 0.67042901657400 & 3.89621469591328 & 0.28658199996032 & 0.33580079990540 & -8.79611297676496  & 2.30107791124568 & 0.5971863 & 0.8863193 & 280.05 & 168.93 \\
 \cmidrule{2-11}
13 & 0.63101207035900 & 0.43965427972022 & -1.86340151687343 & -3.57720228268502 & -0.90767835138654  & 2.18411803952933 & 0.1997748 & 0.5404727 & 162.61 & 109.47 \\
 \cmidrule{2-11}
14 & 0.63457627583700 & 0.10193714593901 & -2.17451765594779 & -4.40347076044376 & -0.24683242373353  & 2.22618108772643 & 0.2676723 & 0.5874529 & 168.38 & 96.42 \\
 \cmidrule{2-11}
15 & 0.63556300684100 & 0.15692422204782 & -1.81467973423677 & -3.34196530655382 & -0.31953595751770  & 2.12490532903941 & 0.1603514 & 0.5207836 & 171.53 & 98.07 \\
 \cmidrule{2-11}
16 & 0.72887955161100 & -1.24078151598096 & -2.06919285860216 & -3.27703304247117 & 2.04486448137144  & 1.72920803153639 & 0.1590343 & 0.7698991 & 198.62 & 55.24 \\
 \cmidrule{2-11}
17 & 0.69323912398600 & -1.14722098636553 & -1.90205344064335 & -3.20352292114426 & 1.99917787967515  & 1.85996720106556 & 0.1537037 & 0.7203409 & 195.34 & 55.30 \\
 \cmidrule{2-11}
18 & 0.63929871699400 & -2.00041668747959 & 1.31858790683704 & 3.42097602853642 & 4.26683084450913  & 2.37565322811402  & 0.4760875 & 0.9055175 & 304.98 & 307.25\\
 \cmidrule{2-11}
19 & 0.64773137699600 & -1.10753033827687 & -1.51623982336362 & -2.64158658109620 & 2.03671753399746  & 1.98344156055357  & 0.0701832 & 0.7175660 & 201.31 & 48.79\\
 \cmidrule{2-11}
20 & 0.72914027110900 & 1.94120755448657 & 1.12322966108835 & 1.80321403347542 & -2.93121977382635  & 1.64752624143655  & 0.0532524 & 0.8560548 & 156.53 & 213.60\\
 \cmidrule{2-11}
\bottomrule\end{tabular}}\\
{Initial conditions of the families 1-20 of asymmetric periodic orbits in 2/1 MMR in coordinates of the rotating frame ($\dot\theta$ being the angular velocity of the $Ox$-axis with respect to the inertial frame) of ERTBP shown in Figs. \ref{orb1}-\ref{orb4}. We also provide the eccentricities and the angles $\Delta M=M_2-M_1$ and $\Delta\varpi=\varpi_2-\varpi_1$. The orbits start with $\dot x'=0$ and are unstable.}
\label{as}
\end{table*}

\begin{landscape}
\section{Possible locations of long-term stable inner terrestrial planets in 22 single-giant planet systems}
\label{possible}
\begin{table}[!h]
\caption{Possible locations of long-term stable inner terrestrial planets in the 22 single-giant planet systems with $e_2\geq 0.4$.}
\scalebox{0.47}{\begin{tabular}[b]{ccccccccccccccccccccccc}
\toprule
  &&&&&&&&&&&& MMR &&&&&&&&&& \\
 \cmidrule{2-23}
 & $3/2$ & $5/3$ & $2/1$ & $7/3$& $5/2$ & $8/3$ & $3/1$ & $7/2$ & $4/1$ & $9/2$ &$5/1$& $11/2$ &$6/1$& $13/2$ &$7/1$& $15/2$ &$9/1$& $19/2$ &$12/1$& $25/2$ & $19/1$ & $23/1$\\
 \cmidrule{2-23}
 &&&&&&&&&&&& $a_1/a_2$ &&&&&&&&&& \\
 \cmidrule{2-23}
& $0.7631$ & $0.7113$ & $0.6299$ & $0.5684$& $0.5428$ & $0.5200$ & $0.4807$ & $0.4337$ & $0.3968$ & $0.3668$ &$0.3419$& $0.3209$ &$0.3028$& $0.2871$ &$0.2732$& $0.2609$ &$0.2311$& $0.2229$&$0.1907$ & $0.1856$ & $0.1404$ & $0.1236$\\
 \cmidrule{2-23}
${\rm System \; HD}$  &  &&&& &  & &  &  &&&  &&&  &  &  &  &  &  &  &  \\
\midrule
$20782$  & &  & \multirowbt{1}{*}{$e_1\geq0.3$} && &  &\multirowbt{1}{*}{$e_1\geq0.3$} &  & \multirowbt{1}{*}{$(e_1\geq0.3)$} &  &\multirowbt{1}{*}{$(e_1=0.5)$}  &  &\multirowbt{1}{*}{$(e_1=0.5)$}  & &\multirowbt{1}{*}{$(e_1=0.5)$} &&\multirowbt{1}{*}{$(e_1=0.5)$}&&\multirowbt{1}{*}{$(e_1=0.5)$}&&&   \\
 \cmidrule{1-23}
$108341$  & &  & \multirowbt{1}{*}{$e_1\geq0.3$} && &  &\multirowbt{1}{*}{$e_1\geq0.3$} &  & \multirowbt{1}{*}{$(e_1\geq0.3)$} &  &\multirowbt{1}{*}{$e_1=0.5$\rm{\;E}}  &  &\multirowbt{1}{*}{$e_1=0.5$\rm{\;E}}  & &\multirowbt{1}{*}{$e_1=0.5$\rm{\;E}} &&\multirowbt{1}{*}{$(e_1=0.5)$\rm{\;E}}&&\multirowbt{1}{*}{$e_1=0.5$\rm{\;E}}&&&  \\
 \cmidrule{1-23}
$4113$   & &  & \multirowbt{1}{*}{$e_1\geq0.3$} && &  &\multirowbt{1}{*}{$e_1\geq0.3$} &  & \multirowbt{1}{*}{$(e_1\geq0.3)$} &  &  &  &\multirowbt{1}{*}{$(e_1=0.5)$}  &&&  &&&&&& \\
 \cmidrule{1-23}
$28254$& &  & \multirowbt{1}{*}{$e_1\geq0.3$} & & &&$e_1=0.5$&$e_1=0.02$\rm{\;E} &$e_1=0.5$\rm{\;E}  &$(e_1=0.02)$  &  & & &$(e_1\geq0.02)$& &  &&  & &$e_1=0.3$\rm{\;E} &  & \\
\cmidrule{1-23}
$45350$  & &  & \multirowbt{1}{*}{$e_1\geq0.3$} & & &$(e_1=0.3)$&$e_1=0.5$&$e_1=0.02$ &  &  & & && & &$(e_1=0.3)$&  &$(e_1=0.3)$  &  && \\  
 \cmidrule{1-23}
\multirowbt{1}{*}{$30562$}  & &  &\multirowbt{1}{*}{{$e_1\geq0.3$}}  &\multirowbt{1}{*}{{$(e_1=0.3)$}}& \multirowbt{1}{*}{$e_1=0.02$} & &\multirowbt{1}{*}{$e_1=0.5$} &\multirowbt{1}{*}{$(e_1=0.1)$\rm{\;E}}  & &\multirowbt{1}{*}{$(e_1=0.3)$\rm{\;E}}& &\multirowbt{1}{*}{ $(e_1=0.3)$\rm{\;E}}&& \multirowbt{1}{*}{$(e_1=0.1)$\rm{\;E}} && $(e_1=0.1,0.3)$\rm{\;E}& &\multirowbt{1}{*}{$(e_1=0.3)$\rm{\;E}} &&\multirowbt{1}{*}{ $(e_1=0.3)$\rm{\;E}} &\multirowbt{1}{*}{$(e_1=0.3)$\rm{\;E}}& \multirowbt{1}{*}{$(e_1=0.3)$\rm{\;E}} \\
 \cmidrule{1-23}
$86226$   & \multirowbt{2}{*}{$e_1\geq0.02$\rm{\;E}} & \multirowbt{2}{*}{$e_1\geq0.02$\rm{\;E}} & \multirowbt{2}{*}{$e_1=0.5$} & \multirowbt{2}{*}{$e_1\geq0.02$\rm{\;E}}& \multirowbt{2}{*}{$e_1\geq0.02$} & \multirowbt{2}{*}{$(e_1\geq0.02)$\rm{\;E}} & \multirowbt{1}{*}{$(e_1=0.3)$\rm{\;E}} &\multirowbt{2}{*}{$(e_1=0.1)$\rm{\;E}}  &\multirowbt{1}{*}{$(e_1=0.5)$\rm{\;E}} &\multirowbt{2}{*}{$(e_1=0.3)$\rm{\;E}}&\multirowbt{1}{*}{$(e_1=0.5)$\rm{\;E}} &\multirowbt{2}{*}{ $(e_1=0.3)$\rm{\;E}}&\multirowbt{1}{*}{$(e_1=0.5)$\rm{\;E}}& \multirowbt{2}{*}{$(e_1=0.1)$\rm{\;E}} &\multirowbt{1}{*}{$(e_1=0.5)$\rm{\;E}}& \multirowbt{2}{*}{$(e_1=0.3)$\rm{\;E}}&\multirowbt{1}{*}{$(e_1=0.5)$\rm{\;E}} &\multirowbt{2}{*}{$(e_1=0.3)$\rm{\;E}} &\multirowbt{1}{*}{$(e_1=0.5)$\rm{\;E}}&\multirowbt{1}{*}{ $(e_1=0.1)$\rm{\;E}} &\multirowbt{2}{*}{$(e_1=0.3)$\rm{\;E}}& \multirowbt{2}{*}{$(e_1=0.3)$\rm{\;E}} \\
\cmidrule{1-1}
$129445$   &  &  &  & &  &  & \multirowbt{1}{*}{$e_1=0.5$\rm{\;E}} &  & && &&& &&& &  &\multirowbt{1}{*}{$(e_1=0.1)$\rm{\;E}}&\multirowbt{1}{*}{$(e_1=0.3)$\rm{\;E}} && \\
 \cmidrule{1-23}
$120084$   & \multirowbt{2}{*}{$e_1\geq0.02$\rm{\;E}} & \multirowbt{2}{*}{$e_1\geq0.02$\rm{\;E}} & \multirowbt{2}{*}{$e_1=0.5$} & \multirowbt{2}{*}{$e_1\geq0.02$\rm{\;E}}& \multirowbt{2}{*}{$e_1\geq0.02$} & \multirowbt{2}{*}{$(e_1\geq0.02)$\rm{\;E}} & \multirowbt{2}{*}{$e_1=0.5$\rm{\;E}} &\multirowbt{2}{*}{$(e_1=0.3)$\rm{\;E}}  & &\multirowbt{2}{*}{$(e_1=0.3)$\rm{\;E}}&&\multirowbt{2}{*}{ $(e_1=0.3)$\rm{\;E}}&& \multirowbt{2}{*}{$(e_1=0.3)$\rm{\;E}} && \multirowbt{2}{*}{$(e_1=0.3)$\rm{\;E}}&& \multirowbt{2}{*}{$(e_1=0.3)$\rm{\;E}} &&\multirowbt{2}{*}{ $(e_1=0.1)$\rm{\;E}} &\multirowbt{2}{*}{$(e_1=0.3)$\rm{\;E}}&    \\
\cmidrule{1-1}
$16175$ & & &  & &  &&&&  &  &  &  &&&  & &  &  &  &  &  & \\
\cmidrule{1-23}
$152079$   & \multirowbt{1}{*}{$e_1\geq0.02$\rm{\;E}} & \multirowbt{1}{*}{$e_1\geq0.02$\rm{\;E}} & \multirowbt{1}{*}{$e_1=0.5$} & \multirowbt{1}{*}{$e_1\geq0.02$\rm{\;E}}& \multirowbt{1}{*}{$e_1\geq0.02$} & \multirowbt{1}{*}{$(e_1\geq0.02)$\rm{\;E}} & \multirowbt{1}{*}{$(e_1=0.3)$\rm{\;E}} &\multirowbt{1}{*}{$(e_1=0.3)$\rm{\;E}}  &\multirowbt{1}{*}{$e_1=0.5$\rm{\;E}} &\multirowbt{1}{*}{$(e_1=0.3)$\rm{\;E}}&\multirowbt{1}{*}{$e_1=0.5$\rm{\;E}} &\multirowbt{1}{*}{ $(e_1=0.3)$\rm{\;E}}&& \multirowbt{1}{*}{$(e_1=0.3)$\rm{\;E}} & &\multirowbt{1}{*}{$(e_1=0.3)$\rm{\;E}}&& \multirowbt{1}{*}{$(e_1=0.3)$\rm{\;E}} &&\multirowbt{1}{*}{ $(e_1=0.1)$\rm{\;E}} &\multirowbt{1}{*}{$(e_1=0.3)$\rm{\;E}}& \\
 \cmidrule{1-23}
$171028$   & \multirowbt{2}{*}{$e_1\geq0.02$\rm{\;E}} & \multirowbt{2}{*}{$e_1\geq0.02$\rm{\;E}} & \multirowbt{2}{*}{$e_1=0.5$\rm{\;E}} & \multirowbt{2}{*}{$e_1\geq0.02$\rm{\;E}}& \multirowbt{2}{*}{$e_1\geq0.02$} & \multirowbt{2}{*}{$(e_1\geq0.02)$\rm{\;E}} &  &\multirowbt{2}{*}{$(e_1=0.3)$\rm{\;E}}  & &\multirowbt{2}{*}{$(e_1=0.3)$\rm{\;E}}&&\multirowbt{2}{*}{ $(e_1=0.3)$\rm{\;E}}&& \multirowbt{2}{*}{$(e_1=0.3)$\rm{\;E}} && \multirowbt{2}{*}{$(e_1=0.3)$\rm{\;E}}& &\multirowbt{2}{*}{$(e_1=0.3)$\rm{\;E}} &&\multirowbt{2}{*}{ $(e_1=0.1)$\rm{\;E}} &\multirowbt{2}{*}{$(e_1=0.3)$\rm{\;E}}&  \\
\cmidrule{1-1}
$79498$   & & &  & &  &  &&&&  &  &  &  &&& &  &  &  &  &  & \\
 \cmidrule{1-23}
$220773$   & \multirowbt{9}{*}{$e_1\geq0.02$\rm{\;E}} & \multirowbt{9}{*}{$e_1\geq0.02$\rm{\;E}} &  & \multirowbt{9}{*}{$e_1\geq0.02$\rm{\;E}}& \multirowbt{9}{*}{$e_1\geq0.02$} & \multirowbt{9}{*}{$(e_1\geq0.02)$\rm{\;E}} &  &\multirowbt{9}{*}{$(e_1\geq 0.02)$\rm{\;E}}  & &\multirowbt{9}{*}{$(e_1\geq0.02)$\rm{\;E}}&\multirowbt{9}{*}{$e_1=0.5$\rm{\;E}} &\multirowbt{9}{*}{ $(e_1\geq0.02)$\rm{\;E}}&& \multirowbt{9}{*}{$(e_1\geq0.02)$\rm{\;E}} && \multirowbt{9}{*}{$(e_1\geq0.02)$\rm{\;E}}&& \multirowbt{9}{*}{$(e_\geq0.02)$\rm{\;E}} &&\multirowbt{9}{*}{ $(e_1\geq0.02)$\rm{\;E}} &\multirowbt{9}{*}{$(e_1\geq0.02)$\rm{\;E}}&  \\
\cmidrule{1-1}
$142415$   & & &  & &  &&&&  &  &  &  &&&  & &  &  &  &  &  & \\
\cmidrule{1-1}
$29021$   & & &  & &  &  & &&& &  &  &  &&& &  &  &  &  &  & \\
\cmidrule{1-1}
$210277$   & & &  & &  &  &  &&&&  &  &  &&& &  &  &  &  &  & \\
\cmidrule{1-1}
$66428$   & & &  & &  &  &  &  & &&& &  & & && &  &  &  &  & \\
\cmidrule{1-1}
$213240$   & & &  & &  &  &  &  &  &&&&  & &  &&&  &  &  &  & \\
\cmidrule{1-1}
$23127$   & & &  & &  &  &  &  &  &  & &&&&  &  &&&  &  &  & \\
\cmidrule{1-1}
$162004$   & & &  & &  &  &  &  &  &  & & &&& &  &&&  &  &  & \\
\cmidrule{1-1}
$171238$   & & &  & &  &  &  &  &  &  & &  & &&& &  &&&  &  & \\
\bottomrule\end{tabular}}
{Possible locations of long-term stable inner terrestrial planets in the 22 single-giant planet systems with $e_2\geq 0.4$. As in Fig. \ref{confall}, whenever a survival of a terrestrial is favoured, we note the minimum value of $e_1$ (among 0.02, 0.1, 0.3, and 0.5) in the MMR's column. The symbol $E$ refers to additional possible captures in MMRs allowed by the observational errors in $e_2$. When the survival is not probable, because the regular domain is not broad or the giant is located at its borders, the value $e_1$ is displayed in parentheses. We note that only the main MMRs are reported here (other tongues exist; they are narrower than those already mentioned and survival is far more improbable for them).}
\label{tab}
\end{table}
\end{landscape}
\vfill
\clearpage

\end{appendix}

\end{document}